\documentclass[%
 aip,
 amsmath,amssymb,
 reprint,%
]{revtex4-1}
\usepackage[utf8]{inputenc} % For UTF-8 encoding
\usepackage[T1]{fontenc}    % For better font encoding
\usepackage{mathptmx}       % Times New Roman font
\usepackage{graphicx}       % For figures
\usepackage{dcolumn}        % Align table columns on decimal points
\usepackage{multirow}
\usepackage{bm}             % Bold math symbols
\usepackage{etoolbox}       % For email formatting
\usepackage{graphicx}% Include figure files
\usepackage{dcolumn}% Align table columns on decimal point
\usepackage{bm}% bold math
\makeatletter
\def\@email#1#2{%
 \endgroup
 \patchcmd{\titleblock@produce}
  {\frontmatter@RRAPformat}
  {\frontmatter@RRAPformat{\produce@RRAP{*#1\href{mailto:#2}{#2}}}\frontmatter@RRAPformat}
  {}{}
}%
\makeatother
\begin{document}

\preprint{AIP/123-QED}

\title[Optimizing CMOS-Compatible, Superconducting Titanium Nitride Resonators:\\Deposition Conditions and Structuring Processes]{Optimizing CMOS-Compatible, Superconducting Titanium Nitride Resonators: Deposition Conditions and Structuring Processes}
% Force line breaks with \\
% Gleichberechtigte Erstautoren
\author{S. J. K. Lang*}
\email{simon.lang@emft.fraunhofer.de}
\affiliation{Fraunhofer Institute for Electronic Microsystems and Solid State Technologies EMFT, Munich, Germany}

\author{A. Schewski*}
\email{alexandra.schewski@emft.fraunhofer.de}
\affiliation{Fraunhofer Institute for Electronic Microsystems and Solid State Technologies EMFT, Munich, Germany}

% Weitere Autoren
\author{I. Eisele}
\affiliation{Fraunhofer Institute for Electronic Microsystems and Solid State Technologies EMFT, Munich, Germany}
\affiliation{Center Integrated Sensor Systems (SENS), Universität der Bundeswehr München, Munich, Germany}

\author{J. Weber}
\affiliation{Fraunhofer Institute for Electronic Microsystems and Solid State Technologies EMFT, Munich, Germany}

\author{C. Moran Guizan}
\affiliation{Fraunhofer Institute for Electronic Microsystems and Solid State Technologies EMFT, Munich, Germany}

\author{Z. Luo}
\affiliation{Technical University of Munich, TUM School of Computation, Information and Technology, Department of Electrical Engineering, Garching, Germany}

\author{M. Singer}
\affiliation{Technical University of Munich, TUM School of Computation, Information and Technology, Department of Electrical Engineering, Garching, Germany}

\author{B. Schoof}
\affiliation{Technical University of Munich, TUM School of Computation, Information and Technology, Department of Electrical Engineering, Garching, Germany}

\author{M. Tornow}
\affiliation{Fraunhofer Institute for Electronic Microsystems and Solid State Technologies EMFT, Munich, Germany}
\affiliation{Technical University of Munich, TUM School of Computation, Information and Technology, Department of Electrical Engineering, Garching, Germany}

\author{T. Mayer}
\affiliation{Fraunhofer Institute for Electronic Microsystems and Solid State Technologies EMFT, Munich, Germany}

\author{D. Zahn}
\affiliation{Fraunhofer Institute for Electronic Microsystems and Solid State Technologies EMFT, Munich, Germany}

\author{R. N. Pereira}
\affiliation{Fraunhofer Institute for Electronic Microsystems and Solid State Technologies EMFT, Munich, Germany}

\author{C. Kutter}
\affiliation{Fraunhofer Institute for Electronic Microsystems and Solid State Technologies EMFT, Munich, Germany}
\affiliation{Center Integrated Sensor Systems (SENS), Universität der Bundeswehr München, Munich, Germany}
\collaboration{*These authors contributed equally to this work and are listed in alphabetical order.}

\date{\today}% It is always \today, today,
\begin{abstract}
We report on the fabrication and characterization of superconducting coplanar waveguide (CPW) resonators based on titanium nitride (TiN) thin films deposited on 200\,mm diameter high-resistivity Si(100) substrates. We systematically investigate how deposition conditions, dry-etch power and in-situ resist strip temperature affect morphology, superconducting properties and dielectric losses. By tuning reactive sputtering conditions, three distinct preferred out-of-plane crystal orientations — (111), (200), and a mix of both are achieved. Our results demonstrate that all films exhibit similar minimal two-level system (TLS) losses, with TiN111 exhibiting the lowest median TLS losses $\tilde{\delta}_\mathrm{TLS}$, and greater robustness against reoxidation. The applied structuring process, in contrast, has a far greater influence on the TLS loss than the crystal orientation of the TiN film and, consequently, the intrinsic material properties of the superconducting layer. The lowest TLS losses for all TiN depositions were achieved with a low power etch and low temperature resist strip. An additional buffered oxide etch (BOE) treatment could remove high-loss interfacial oxides at the metal-air (MA) and substrate-air (SA) interface and recover the etch-induced TLS losses. Consequently, TiN resonators exhibiting $\tilde{\delta}_\mathrm{TLS}$ values as low as $9.67 \times 10^{-7}$ were realized. The corresponding median low-power loss, $\tilde{\delta}_\mathrm{LP}$, amounts to $11.04 \times 10^{-7}$, which translates to an internal quality factor approaching one million. These findings highlight the critical role of process induced oxide formation at the MA and SA interfaces in limiting the performance of TiN resonators and provide a scalable, low-loss process compatible with industry-grade 200\,mm CMOS qubit fabrication workflows.
\end{abstract}
\maketitle

\section{Introduction}

Quantum computing promises to tackle computational tasks that lie beyond the reach of classical machines operating within a practical timeframe. For example, Shor’s algorithm enables efficient factorization of large integers into primes \cite{fowler_Surface_2012}, and Grover’s algorithm facilitates the rapid search of unstructured databases \cite{Grover_Quantum_1997}.
One of the most promising hardware platforms for realizing large-scale quantum processors is based on superconducting qubits. These qubits are lithographically-defined superconducting circuits typically fabricated from aluminium, titanium nitride (TiN), as well as niobium on silicon (Si) or sapphire substrates. The use of Al and TiN enables compatibility with CMOS-style fabrication tools which supports scaling to large qubit counts \cite{Yost_Solid_2020, Arute_Quantum_2019}.
However, executing genuinely useful quantum algorithms in multi-qubit systems requires exceedingly low error rates. When the error rates for single- and two-qubit gates are reduced to the order of about 0.1\% per operation, fault-tolerant quantum error-correction schemes become feasible—enabling logical qubits to be encoded across many physical qubits, and thereby allowing significantly more complex quantum circuits to execute reliably \cite{fowler_Surface_2012}.
A persistent challenge in superconducting platforms is that, at low powers, both qubit coherence and quality of microwave resonators are often limited by energy loss into parasitic two-level systems (TLS) residing at interfaces or in bulk dielectric materials \cite{gao_experimental_2008}. 
Coplanar waveguide (CPW) resonators provide an ideal platform for studying these TLS. When combined with electromagnetic participation-ratio models, their quality factors allow quantitative links between changes in the fabrication process and distinct loss channels, even in cases where fabrication variations couple to multiple loss mechanisms simultaneously.
\\
Titanium nitride (TiN) is a key material for such resonators. 
%In thin films, the elevated kinetic inductance lowers the resonator frequency, which enables a shorter resonator length and thus minimizes the device footprint at least to some extent. 
It has excellent superconducting properties, it is thermally and mechanically robust and compatible with standard CMOS-style fabrication processes, making it especially well-suited for large-scale manufacturing.
Although TiN resonators have been extensively studied in prior work \cite{schoof_development_2024_2, sandberg_etch_2012, vissers_low_2010, ohya_room_2014,chang_improved_2013, calusine_analysis_2018,amin_loss_2022}, a remaining gap lies in the systematic assessment and demonstration of how individual fabrication steps affect resonator quality factors for realistic geometries. In this work we aim to address this gap by analyzing how various process parameters influence resonator performance. Importantly, we use identical resonator geometries across all processing conditions so that the participation ratios remain constant and any changes in quality factor can be attributed to processing rather than geometric variation.
\\
The processes under detailed analysis are:
\begin{itemize}
    \item TiN sputtering with reproducible films of different crystal orientation.
    \item Dry etching of resonators at high and low power.
    \item In-situ resist stripping at high and low power.
    \item Wet etching as a post processing surface etching technique.
\end{itemize}
\section{Methods}

\subsection{Chip Layout}
\label{subsec:Chip_Layout}
\begin{figure}[b]
    \includegraphics[width=\linewidth]{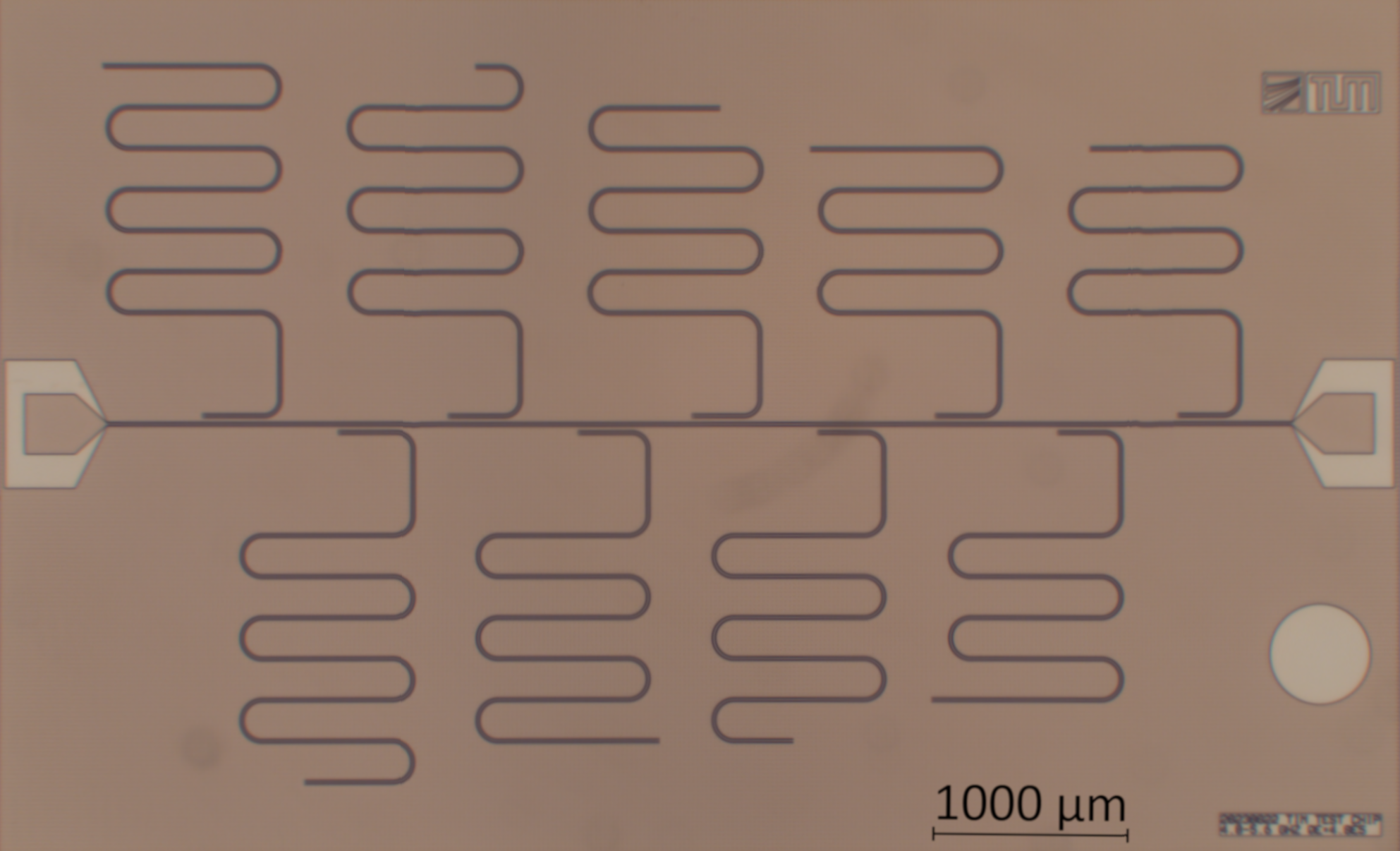}
    \caption{\label{fig:Layout}Exemplary optical microscope image of chip fabricated and characterized within this study.}
\end{figure}

The chip layout used throughout this study can be seen in Figure~\ref{fig:Layout}.
Each chip incorporates nine quarter-wave ($\lambda$/4) resonators, all coupled to a single common feedline for simultaneous signal input and extraction within a single frequency sweep. The coupling to the feedline was designed to be 0.36\,MHz (Q$_\text{ext}$\,=\,5e5) for all of the resonators. The resonators themselves are implemented using a coplanar waveguide (CPW) geometry, featuring a central conductor width of $w=10\,\mu$m and a gap to the ground planes of $g=6\,\mu$m. On each chip, the resonators are spaced 200\,MHz apart in frequency. To cover the full frequency range of 4–10.4\,GHz, four distinct chips were designed, each spanning a specific subset (4–5.6\,GHz, 5.6–7.2\,GHz, 7.2–8.8\,GHz and 8.8–10.4\,GHz) of the range. 
The design also includes vortex-pinning holes around the CPW lines to minimize flux trapping \cite{martinis_superconducting_2009}. 

\subsection{Fabrication}
For the fabrication of the chips whose layout is described in Section~\ref{subsec:Chip_Layout} p-doped Si(100) wafers from Si-Mat with 200\,mm diameter and a resistivity of $>$\,3000\,$\Omega$cm are used as substrates. A 180\,s long clean in 1\% diluted hydrofluoric acid (corrosive GHS05/toxic GHS06 etchant) is used to remove the native oxide and precondition the silicon surface. The wafers are then transferred within 15 minutes to the planar magnetron DC pulsed sputter cluster tool Clusterline 200 from Evatec with a base pressure of $1\cdot10^{-8}$\,mbar. The Ti target has a diameter of 300\,mm, is parallel oriented and centered over the substrate with a distance of 50\,mm. The additional magnetron configuration improves the deposition uniformity. After a preheating step a layer of 150\,nm of TiN is reactively sputtered on the Si surface using three different reactive sputtering processes. The parameter sets for the three tested conditions are summarized in Table~\ref{tab:TiN_process_parameters_2}. They are a subset of process parameters variations tested in an earlier unpublished study and are chosen such that the resulting three films cover a wide range of different film properties in terms of crystallography and texture as will be discussed in detail in Sec~\ref{sec:TiN_character}.
\\
Deposition is followed by SPA-3000 i4 Canon stepper photolithography. 
\begin{table}[b]
\caption{\label{tab:TiN_process_parameters_2}Parameters of reactive sputter processes tested for TiN film deposition.}
\begin{ruledtabular}
\begin{tabular}{ccccc}
\textbf{Depo} & \textbf{Ar [sccm]} & \textbf{N\textsubscript{2} [sccm]} & \textbf{Power [W]} & \textbf{Temp. [$^\circ$C]} \\ 
\hline
A   & 12  & 3.5   & 500  & 30   \\
B   & 12  & 24    & 3000 & 400  \\
C   & 16  & 50    & 6800 & 400  \\		
\end{tabular}
\end{ruledtabular}
\end{table}
For structuring of the TiN resonators a MxP metal etch chamber and an ASP strip chamber available on the same Applied Materials P5000 mainframe are used. 
%advanced strip passivation = ASP
Two different dry etching recipes are tested whose parameters can be found in Table~\ref{tab:TiN_etch_parameters_2} and who are referred to as \textit{high power} (HP) and \textit{low power} (LP) etches.
\\
Following the etching process, the photoresist is removed in-situ using an H$_2$O plasma strip. 
\begin{table}
\caption{\label{tab:TiN_etch_parameters_2}Parameters of TiN dry etch process (Power in Watts, flow rates in sccm, magnetic field in Gauss).}
\begin{ruledtabular}
\begin{tabular}{cccccc}
\textbf{Etch} & \textbf{Power [W]} & \textbf{BCl\textsubscript{3} [sccm]} & \textbf{Cl\textsubscript{2} [sccm]} & \textbf{N\textsubscript{2} [sccm]} & \textbf{B [Gauss]} \\ 
\hline
HP  & 500    & 50  & 20 & 50 & 30  \\
LP  & 70     & 0   & 80 & 40 & 40  \\
\end{tabular}
\end{ruledtabular}
\end{table}
Two different temperatures are tested during this strip process: 250\,$^\circ$C (high temperature, HT) and 150\,$^\circ$C (low temperature, LT). Lastly, all samples undergo a chemical cleaning using CA25 from TechniClean (health hazard GHS07, flammable GHS02). 
As further characterization is conducted at the chip level, the fabrication process at the wafer level was concluded by dicing. 
\\
To investigate the effects of surface oxidation, an additional surface treatment — a wet etching step using Honeywell ammonium fluoride etchant AF 875-125 LST (referred to as BOE; corrosive GHS05, toxic GHS06 HF-based etchant) — was tested at the chip level.  To minimize reoxidation after this cleaning step one of the chips was cooled down within the next two hours (time coupling, tc) for cryogenic characterization. All the other chips were also cooled down within 24 hours after the etching or BOE post treatment.
\\
Table~\ref{tab:process_steps} summarizes all the combined process variations that will be analyzed within the course of this work. The last column also shows how many resonators are used for the characterization.
\begin{table}
\caption{\label{tab:process_steps}Summary of all variations tested within the fabrication flow.}
\begin{ruledtabular}
\begin{tabular}{ccccc}
\textbf{Depo} & \textbf{Etch} & \textbf{Resist strip} & \textbf{Wet etch} & \textbf{No. of Resonators}\\ 
\hline
\multirow{4}{*}{A} 
    & \multirow{2}{*}{HP} & HT & - &17\\
    &                     & LT & - &18\\
    \cline{2-5}
    & \multirow{2}{*}{LP} & HT & - &18\\
    &                     & LT & - &17\\
\hline\hline
\multirow{4}{*}{B} 
    & \multirow{2}{*}{HP} & HT & BOE (with(out) tc) &16\\
    &                     & -  & - &0\\
    \cline{2-5}
    & \multirow{2}{*}{LP} & -  & - &0\\
    &                     & LT & BOE (without tc)&54\\
\hline\hline
\multirow{4}{*}{C} 
    & \multirow{2}{*}{HP} & HT & - &15\\
    &                     & LT & - &18\\
    \cline{2-5}
    & \multirow{2}{*}{LP} & HT & - &9\\
    &                     & LT & - &18\\
\hline
\end{tabular}
\end{ruledtabular}
\end{table}
\subsection{Room temperature characterization}
Numerous room-temperature characterization methods are employed to monitor and evaluate the fabrication process. The uniformity and reproducibility of the TiN thin film deposition are assessed using wafer-scale sheet resistance measurements at nine points across the wafer, performed with a four-point probe test Omnimap instrument. The crystallinity and texture of the TiN thin film are analyzed using a $2\theta/\omega$ X-ray Diffraction (XRD) scan with a Rigaku MiniFlex. Pseudo-Voigt peaks—a linear combination of Gaussian and Lorentzian functions—are fitted to the peaks in the $2\theta/\omega$ scan to determine the peak positions and full width at half maximum (FWHM). The peak positions provide insights into the preferred crystal orientation, while the FWHM offers information about grain sizes when the grain size is smaller than the instrumental resolution \cite{harrington_back_2021}.
A Helios 650 scanning electron microscope (SEM) is utilized to characterize the TiN thin films in terms of layer thickness and morphology. After structuring, the etching rim and the overetch depth into the substrate are also examined using SEM.
Atomic force microscopy (AFM) measurements, performed with a Park Systems NX20, are used to evaluate the influence of the etching process on the substrate roughness by comparing the Si surface before TiN deposition and after etching.
Finally, the oxidation of the Si surface is monitored using a spectral ellipsometer from SENTECH Instruments.

\subsection{Cryogenic characterization}
\begin{figure}
    \includegraphics[width=\linewidth]{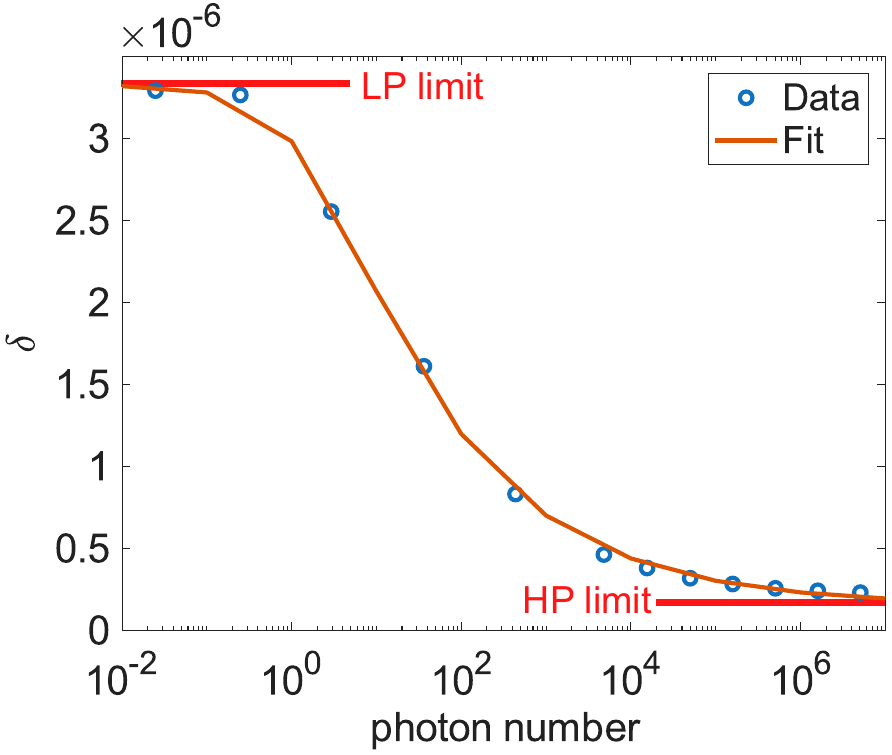}
    \caption{\label{fig:resonator_limits}Exemplary fit of Eq. (3) (orange) on the dielectric loss (blue) extracted of one resonator measured at $T<25\,$mK. The LP and HP limit are marked in red.}
\end{figure}

To determine the electrical and superconducting properties, the TiN thin films are wire-bonded, and temperature-dependent four-terminal measurements are carried out using a Kiutra system. The precise temperature sweep allows the determination of the critical temperature ($T_c$).
The residual resistance ratio (RRR), defined as
\begin{eqnarray}
    RRR_\rho = \frac{\rho(300\,\text{K})}{\rho_0},
\end{eqnarray}
can be determined from the ratio of the resistivity at room temperature, $\rho(300\,\text{K})$, where electron-phonon scattering dominates, and the residual resistivity $\rho_0$, which arises solely from impurities and defects. The latter is temperature-independent and can be measured at temperatures approaching $0\,\text{K}$, where the temperature-dependent behavior of the resistivity transitions from a $\rho \propto T^5$ dependence to a constant offset.
Higher RRR values therefore correspond to lower defect densities \cite{Gross_Festkörperphysik_2018}.
Using only one wire bond configuration and assuming a temperature-independent electric field distribution for this wire bond configuration, $RRR_\rho$ is approximated in this work by
\begin{eqnarray}
    RRR_\rho \approx RRR = \frac{R(300\,\text{K})}{R(\text{just above } T_c)}.
\end{eqnarray}
The denominator is taken as $R(\text{just above } T_c)$, as is commonly done for superconductors, since the resistance vanishes below $T_c$. This approximation assumes that phonon contributions are already negligible in this temperature range.
\\
The cryogenic characterization of the resonators is done in a Bluefors dilution refrigerator and a cryostat from Qinu at temperatures below T$\,<\,25\,$mK.
The measurement process begins with identifying the resonance peaks through a coarse transmission sweep. Following this, the complex scattering parameter $S_{21}$ is measured using a vector network analyzer (VNA) \cite{simons_coplanar_2001, goppl_coplanar_2008}, which is designed to generate and analyze microwave signals. The internal quality factor $Q_{\rm i}$ is determined by fitting the measured data using the procedure outlined in Ref.~\onlinecite{probst_efficient_2015}. 
The excitation power of the VNA can be converted into the corresponding photon number $\langle n \rangle$ following the method described by A. Bruno et al. in Ref.~\onlinecite{bruno_surface_2015}.
The internal quality factor is inversely related to the dielectric loss ($Q_{\rm i} = 1/\delta$).
TLS defects couple to the electromagnetic fields of the resonator and introduce dissipation and dephasing \cite{muller_towards_2019}.
For high powers ($\delta_{\rm HP}$) the dielectric loss decreases as the TLS get saturated. For low powers ($\delta_{\rm LP}$) the TLS loss dominates such that $\delta_{ \rm TLS}$ can be extracted by the following simplified equation \cite{martinis_decoherence_2005}: 
\begin{eqnarray}
    \delta_{\rm TLS} = \delta_{\rm LP} - \delta_{\rm HP}
\end{eqnarray}
One resulting exemplary measurement of the material losses depending on $\langle n \rangle$ is shown in Figure~\ref{fig:resonator_limits}. 
The data points describe a curve given by
\begin{eqnarray}
\delta(\langle n\rangle) 
= \delta_{\rm TLS}\frac{1}{(1 + \langle n\rangle / n_{\rm c})^\beta}
+ \delta_{\rm HP},
\end{eqnarray}
under the assumption of a broad distribution of TLS in the continuum approximation and in the limit $T \to 0$. Here, $n_{\rm c}$ is the critical photon number and $\beta$ is a fit parameter that characterizes how rapidly the TLS contribution decreases with
increasing drive (see reference \cite{martinis_decoherence_2005} and related TLS models).
\\
Each resonator is analyzed individually, and the variation in $\delta_\mathrm{TLS}$ and $\delta_\mathrm{LP}$ is statistically evaluated using the interquartile range (IQR) and its median values, $\tilde{\delta}_\mathrm{TLS}$ and $\tilde{\delta}_\mathrm{LP}$ for multiple resonators.

\subsection{Electro-magnetic-participation-ratio model}

\begin{figure}
    \includegraphics[width=\linewidth]{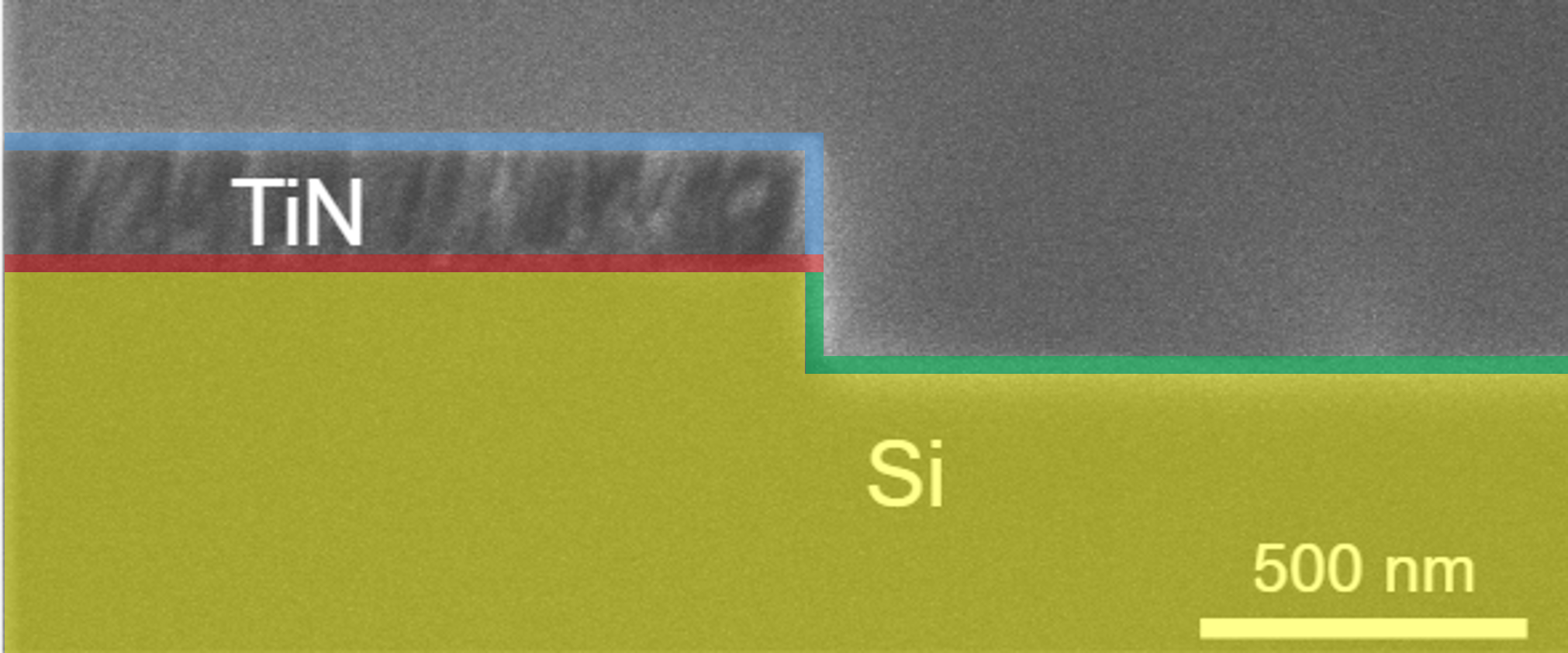}
    \caption{\label{fig:CPW-res}SEM image of CPW resonator in cross sectional view with the dielectric regions contributing to TLS loss marked in yellow (Si), red (MS), blue (MA) and green (SA).}
\end{figure}

For CPW resonators the electro-magnetic-participation-ratio model describes how TLS-related losses can be analyzed \cite{wang_surface_2015, ganjam_surpassing_2024}. 
%check citations!!!
A cross-section of a CPW resonator is shown in Fig.~\ref{fig:CPW-res}, where four dielectric regions $i$ contributing to TLS loss are indicated: the bulk silicon substrate ($i= Si$, yellow), the metal–substrate interface ($i=MS$, red), the metal–air interface ($i=MA$, blue), and the substrate–air interface ($i=SA$, green). Each region has its own loss tangent $\delta_{\rm i}$.
Depending on the resonator design, the so‐called participation ratio \(p_{\rm i}\)—defined as the fraction of the total electromagnetic energy stored (more precisely: the fraction of the energy stored in the electric field \(\vec{\mathfrak{E}}\) in a length element \(s\), divided by the total energy per unit length \(E_{\rm tot}\)) in region \(i\), with thickness \(t_{\rm i}\) and dielectric constant \(\epsilon_{\rm i}\)—can be varied \cite{wenner_surface_2011,gao_experimental_2008}.
The participation ratio $p_{\rm i}$ is then defined by 
\begin{eqnarray}
	p_\text{i} = \frac{t_{\rm i} \epsilon_{\rm i}}{E_\text{tot}} \int d\vec s |\vec{\mathfrak{E}}|^2,
	\label{eq:loss_energy}
\end{eqnarray}
The resulting $\delta_{TLS}$ is given as the sum of all loss tangent $\delta_{\rm i}$ weighted with their respective participation ratio $p_{\rm i}$ according to the following equation:
\begin{eqnarray}
    \delta_{\rm TLS} = p_{\rm MA}\delta_{\rm MA} + p_{\rm MS}\delta_{\rm MS} + p_{\rm SA}\delta_{\rm SA} + p_{\rm Si}\delta_{\rm Si}
\end{eqnarray}
Because the participation ratios depend sensitively on resonator geometry parameters like signal-line width, gap between signal and ground, film thickness and over-etch depth into the silicon substrate one can tailor $p_{\rm i}$ by design. The loss tangent $\delta_{\rm TLS}$ can be determined experimentally by cryogenic measurements.
\\
Disentangling the individual $\delta_{\rm i}$, however, is challenging: typically, the $\delta_{\rm i}$ differ by orders of magnitude such that the influence of some loss region might not be resolved when the resonator is dominated by another loss. In the worst-case scenario, processes such as anisotropic trenching can couple simultaneously to multiple loss channels and, by altering the geometry, also change the participation ratios \cite{woods_determining_2019}.
HFSS simulations summarized in Table~\ref{tab:PPR} show how the participation ratios for loss contributing regions i vary for expected overetch depths of $\leq\,100\,$nm.
\begin{table}
\caption{\label{tab:PPR}Simulated participation ratios of CPW resonators with 10\,$\mu$m width and 6\,$\mu$m gap for different trench depths with assumed dielectric layer thickness of 2\,nm and constants $\varepsilon_{SA}$ = 4, $\varepsilon_{MA}$ = 10, $\varepsilon_{MS}$ = 11.4 and $\varepsilon_{Si}$ = 11.9.}
\begin{ruledtabular}
\begin{tabular}{ccccc}
\textbf{Trench depth [nm]} & \bm$p_\text{SA}$ & \bm$p_\text{MA}$ & \bm$p_\text{MS}$ & \bm$p_\text{Si}$ \\ 
\hline
0   & $2.83 \cdot 10^{-4}$   & $4.95 \cdot 10^{-5}$    & $5.93 \cdot 10^{-4}$    & 0.907 \\
50  & $2.67 \cdot 10^{-4}$   & $2.08 \cdot 10^{-5}$    & $5.45 \cdot 10^{-4}$    & 0.905 \\
100 & $2.51 \cdot 10^{-4}$   & $1.77 \cdot 10^{-5}$    & $5.04 \cdot 10^{-4}$    & 0.903 \\
\end{tabular}
\end{ruledtabular}
\end{table}
\newline
 
\section{Results and Discussion}
\subsection{Room temperature characterization}
\subsubsection{TiN deposition characterization}
\label{sec:TiN_character}

\begin{figure}
    \includegraphics[width=\linewidth]{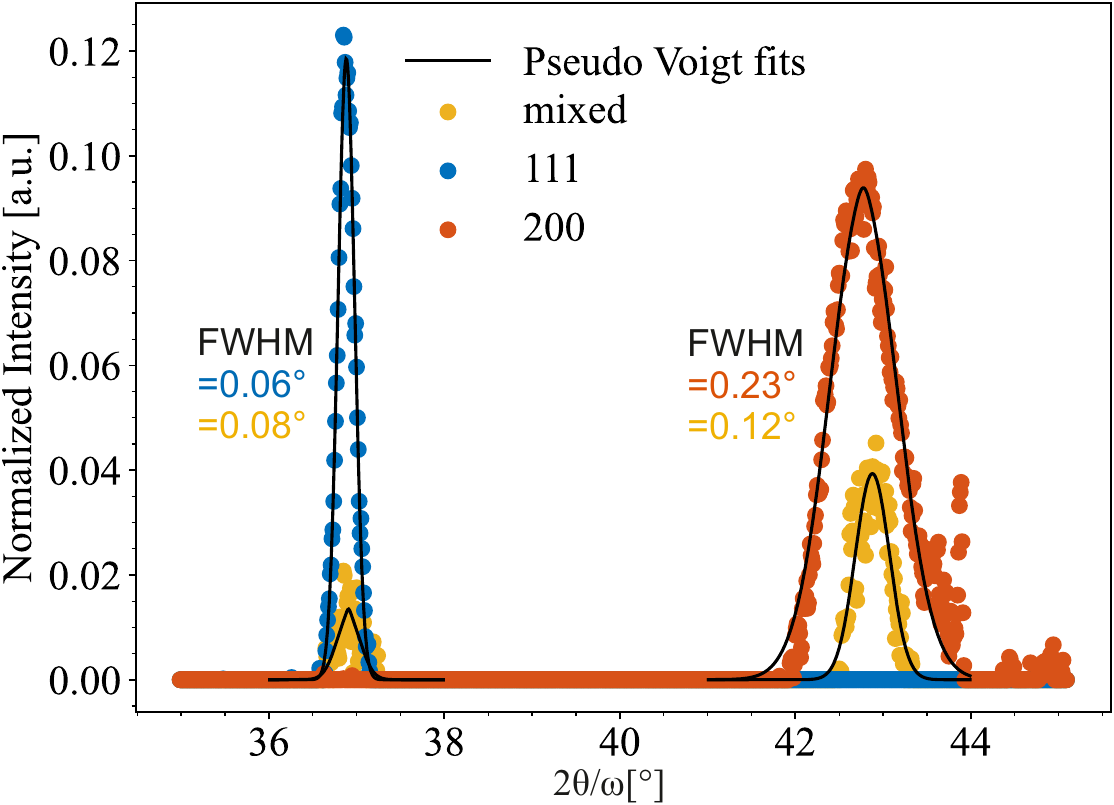}
    \caption{\label{fig:XRD}XRD measurements of the different TiN depositions (A=111, B=mixed, C=200). The peaks at 42.5$^\circ$ and 36.5$^\circ$ correspond to different TiN crystal orientations: (200) and (111), respectively.}
\end{figure}

\begin{figure*}[t]
    \includegraphics[width=\linewidth]{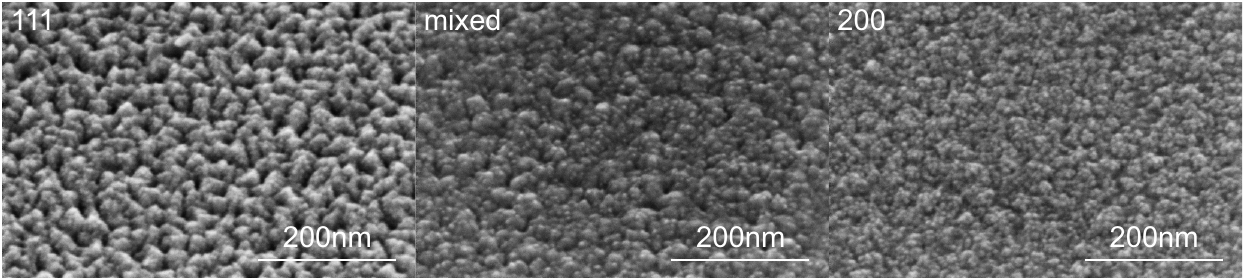}
    \caption{\label{fig:TiNdepoSEM_surface}SEM images of TiN deposited under different conditions to achieve preferential crystal orientations of (111), (200) and a mixture of both.}	
\end{figure*}

For each of the three sputter processes (A, B, and C), four wafers are being deposited. The resulting twelve wafers are characterized using sheet resistance measurements to determine the uniformity and the reproducibility of the process. 
Table~\ref{tab:TiN_results} summarizes the mean of all sheet resistances measured on the batch and
its maximum standard deviation, as well as the maximum
point-wise standard deviation over the entire batch.
\begin{table}
\caption{\label{tab:TiN_results}Sheet resistance statistics on four wafers each for TiN films deposited using the parameters from Table~\ref{tab:TiN_process_parameters_2} and resulting crystal orientations. Be aware that column five shows the maximum \textit{pointwise} $\sigma$ within the batch.}
\begin{ruledtabular}
\begin{tabular}{cccccc}
\textbf{Depo}  & \bm$R_\square$ & \textbf{Max $\sigma$} & \textbf{Max $\sigma$} & \bm$\rho$ & \textbf{Crystal} \\ 
               & \textbf{[$\Omega/sq$]} & \textbf{wafer [\%]} & \textbf{batch [\%]} & \textbf{[$\mu\Omega\cdot$cm]} & \textbf{orientation} \\ 
\hline
A   & 11.9 & 19.0 & 9.7 & 203 & 111 \\ %R = 11.9
B   & 4.35 & 4.25 & 3.6 &  74 & Mixed \\ %R = 4.35
C   & 4.7  & 4.0  & 2.0 &  65 & 200 \\ %R = 4.7
\end{tabular}
\end{ruledtabular}
\end{table}
While process A done at lower temperatures shows a stronger nonuniformity, process B and C have a uniformity below 5\% over one wafer as well as within the whole batch. 
%One possible explanation of the poorer performance especially in terms of uniformity for process A might be the less uniform nitrogogen distribution in the gas phase for reactive sputtering. 
The reduced uniformity may stem from the low power used for process A, which approaches the minimum threshold of the deposition chamber’s capabilities. 
\newline
For each film the thickness d of the TiN layers is measured in a cross-sectional SEM image (like the one visible in Figure~\ref{fig:TiNdepoSEM}) and results together with the corresponding sheet resistance measurement in the resistivity $\rho$ of the film according to $\rho = R_\square \cdot d$. 
While the films B and C have with $\rho = 74\,\mu\Omega\cdot$cm and $\rho = 65\,\mu\Omega\cdot$cm similar resistivities, film A shows a value of $203\,\mu\Omega\cdot$cm that is around three times as large.
We select measurement spots and pick chips from the wafer such that all point‑wise characterizations (XRD, Tc, etc.) of the TiN films, as well as all measured resonators, have the same TiN thickness of about $150\,$nm.
The XRD measurements of all three thin films are shown in Figure~\ref{fig:XRD}. They allow for a detailed characterization of the crystallographic properties like grain size and orientation.
Pseudo-Voigt profiles (indicated in black) are fitted to all of the occurring peaks to determine the peak positions and full width at half maximum (FWHM). The resulting FWHM values are annotated in Figure~\ref{fig:XRD}.
The spectra show peaks at two different positions.
The peaks at 36.9$^\circ$ can be attributed to the TiN(111) \cite{ICDD_00-038-1420_TiN, ohya_sputtered_2013} peak while the peaks at 42.8$^\circ$ are identified to be TiN(200) peaks.
The peaks are slightly shifted to higher angles (0.3$^\circ$ and 0.2$^\circ$), which can be attributed to a slight misalignment of the samples as the Si(400) reference peak also shows this shift.
%The film sputtered using recipe A shows only the TiN(111) peak. Since the $2\theta/\omega$ scan provides information only about out-of-plane lattice parameters, it can be concluded that the (111) orientation is the preferred out-of-plane crystal orientation of this film. Therefore, this film will be referred to as the TiN111 film from now on.
%The thin film resulting from recipe C shows only the TiN(200) peak. 
%Hence, this film has a preferred out-of plane crystal orientation of (200) and will therefore be referred to as the TiN200 film.
%The XRD spectra of the thin film resulting from recipe B exhibits both peaks and therefore the film shows both out-of plane crystal orientations. It will be named TiNmixed from now on.
Recipe A produces a film with only the TiN(111) peak, indicating a (111) out-of-plane orientation (TiN111). Similarly, Recipe C yields only the TiN(200) peak, corresponding to a (200) orientation (TiN200). Recipe B shows both peaks, indicating mixed orientations (TiNmixed).
Here, we can identify a far higher TiN(200) peak compared to the TiN(111), which is varying from the expected intensity ratio of equally distributed grains $I_{111}/I_{200} = 0.72$ \cite{Zoestbergen_Xray_2000, ICDD_00-038-1420_TiN}.
We can therefore conclude that the TiNmixed film has a particularly small amount of (111) oriented grains compared to (200) oriented grains.
\\
Additionally, the peak widths allow conclusions to be drawn about the crystal grain size and micro-strain \cite{harrington_back_2021}, as they are broader than the resolution-limited Si reference peak (FWHM$_{Si}=0.02^\circ$). 
SEM images (see Fig.~\ref{fig:TiNdepoSEM_surface}) reveal fine grains (\textless\,40\,nm) across all depositions with clear difference in grain sizes between the films, suggesting crystal grain size as the main cause of the peak broadening.
In detail, TiN111 films show larger, well-defined grains with pronounced broad grain boundaries, whereas the TiN200 film appears to have smaller grains with less notable grain boundaries. The TiNmixed film resembles the TiN200 film but includes islands that resemble TiN111 grains. 
This aligns with the XRD, where the TiN111 peak has the smallest FWHM, while the TiN200 peak exhibits a FWHM that is nearly five times larger. The TiNmixed film lies in between. Due to the inverse relationship between grain size and FWHM (given by the Scherrer equation where grain size $\propto \frac{1}{\textbf{(FWHM-FWHM}_{Si})\cdot cos(\Theta)}$), the average grains size in the TiN111 film is approximately four times larger than the average grain size in the TiN200 film \cite{harrington_back_2021}.

The XRD analysis together with the SEM images proves that the different deposition parameter sets produced three distinct TiN thin films. This diversity is beneficial for subsequent analysis, as it allows investigation of the influence of process parameters on the resulting dielectric losses across a wide range of film characteristics.
In particular, the influence of grain size and occurrence of grain boundaries and their behavior under etching and oxidation will be revisited again later in this paper.

\subsubsection{TiN structuring characterization}
\label{subsec:struc}

\begin{figure}
    \includegraphics[width=\linewidth]{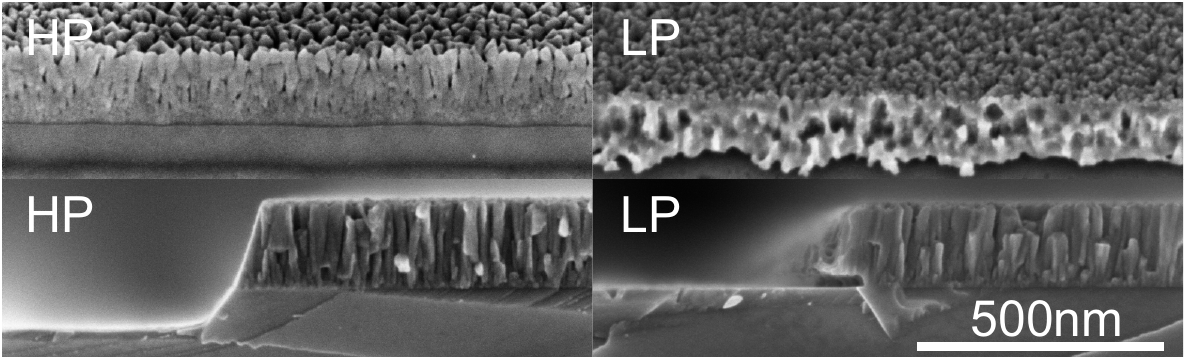}
    \caption{\label{fig:TiNdepoSEM}SEM images of TiN111 etched with HP (left) and LP (right) etch processes in side view (upper images) and cross-sectional view after cleaving (lower images).}
\end{figure}

Figure~\ref{fig:TiNdepoSEM} shows TiN films after two different structuring processes have been applied, whose parameters are described in Table~\ref{tab:TiN_etch_parameters_2}.
The influence of the strip process (here the HT strip) is not visible in the SEM images; its microscopic effects will be discussed below. 
The HP recipe produces steep, well-defined sidewalls. The plan view in the upper image shows a clean etching edge in the TiN and a straight overetch into the silicon. The lower image provides a cross-sectional view of the etching edge after cleaving, enabling measurement of the overetch depth into the silicon. All samples within this work exhibit an overetch depth in the range of 20–100\,nm. In contrast, the LP etch results in sidewalls that appear rough and inhomogeneous, with no visible overetch into silicon, even for extended etch times. Instead, it seems that the TiN near the interface begins to be etched, resulting in a slight underetch.
During the process, the parameters listed in Table~\ref{tab:TiN_etch_parameters_2} result in a DC bias of the susceptor where the wafer is positioned. The HP recipe generates a bias of 180\,V (for 500\,W), while the LP recipe produces a bias of only 17\,V (for 70\,W). This indicates that the kinetic energy of the particles impacting the etched regions is an order of magnitude higher for the HP etch compared to the LP etch. This explains the steep sidewalls observed in Figure~\ref{fig:TiNdepoSEM} and the larger overetch into the silicon. Therefore, the HP etch can be categorized as a process with a strong physical component and finite etch selectivity. The heavier decomposition products of BCl$_3$ are likely contributors to this physical component.
The LP etch, on the other hand, is primarily a chemical etch process, as evidenced by its high selectivity to silicon and the rough appearance of the sidewalls. The roughness of these sidewalls might be attributed to differing chemical etch rates for grain boundaries and TiN grains. 
AFM measurements of the etched silicon surface support the classification into chemically and physically dominated etch regimes. The reference RMS value of a blank silicon wafer is 0.7\,nm. The HP etch increases the surface RMS to 1.2\,nm, while the LP recipe causes only slight roughening, yielding an RMS value of 0.8\,nm.
\\
The resist strip after etching is performed using water vapor plasma at different temperatures (LT and HT). During this step, the exposed silicon surface becomes oxidized. To investigate the oxidation behavior during the strip processes, spectral ellipsometry was used to measure the oxide thickness of three wafers using a Cauchy layer (with floating Cauchy coefficients N0, N1). According to these measurements, an HF-dipped silicon wafer has an oxide thickness of 0.5\,nm. After processing with the LT resist strip, the oxide thickness increases to 1.5\,nm, and after the HT resist strip, it increases further to 3.2\,nm. Although ellipsometer fits 
%(see Fig.~\ref{fig:Elipsometer_Si_HT} and ~\ref{fig:Elipsometer_Si_LT} in the appendix) 
for such small layer thicknesses do not provide reliable absolute values, the increase in oxide thickness with higher temperature is certain. 
%Furthermore, the increase in surface roughness observed in AFM measurements is likely to promote further oxidation.
A similar behavior can be expected for the oxynitride formed on TiN, even though the TiN surface is exposed to oxygen only during the final moments of the resist strip process. TiN oxidizes to rutile TiO$_2$ at temperatures above 600$^\circ$C in the absence of plasma \cite{chen_oxidation_2005}, indicating that the formation of TiO$_2$ is thermodynamically favored over TiN. Since TiN forms an oxynitride layer on its surface even at room temperature, it is reasonable to assume that elevated temperatures and exposure to oxygen plasma during processing further increase the oxynitride thickness. As the resist strip is the final step in the process chain at elevated temperatures, we assume that no further oxidation occurs afterward.
\\
%The following subsection describes how oxidation at the metal–air (MA) and substrate–air (SA) interfaces affects the energy losses in the resonators, and how removing these oxides using buffered oxide etch (BOE) alters those losses. Furthermore, the findings regarding crystal orientation and grain size are incorporated into this discussion, as they directly impact the oxidation behavior.

\subsection{Superconducting properties and material losses}
\label{subsection:resonator_loss}
%(see Sec.~\ref{section:resonator_theory})
\subsubsection{Influence of TiN material}
\begin{figure}
    \includegraphics[width=\linewidth]{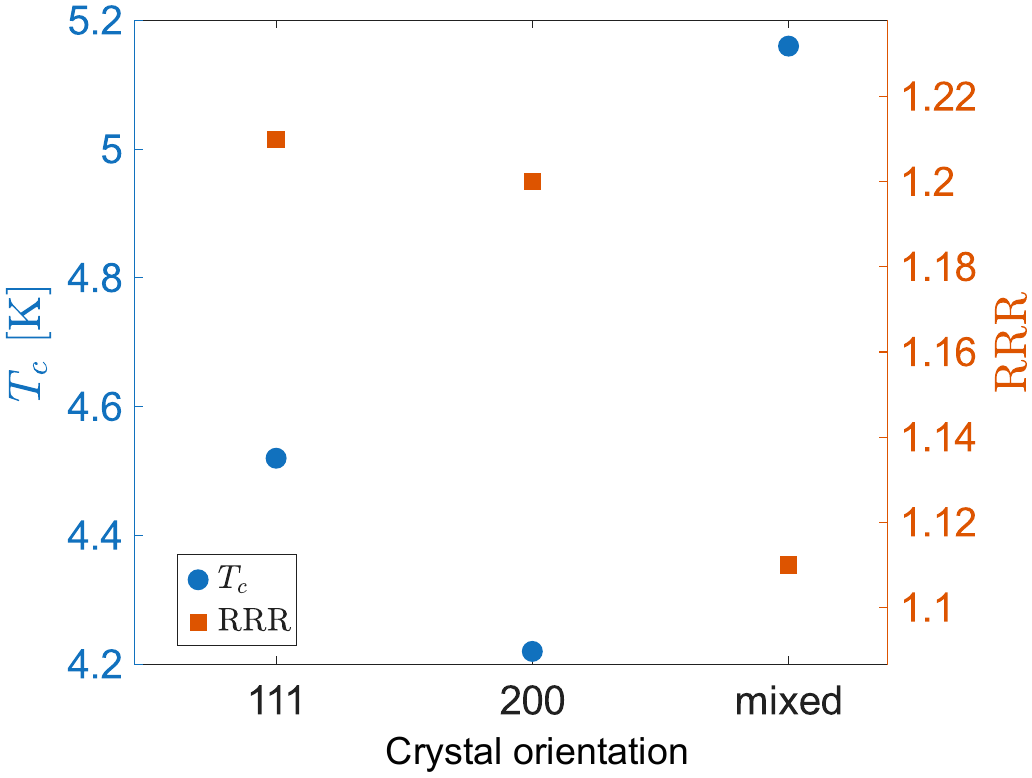}
    \caption{\label{fig:TcRRR}Critical temperature (\( T_c \)) and RRR measured on films with different crystal orientations.}
\end{figure}

\begin{figure}
    \includegraphics[width=\linewidth]{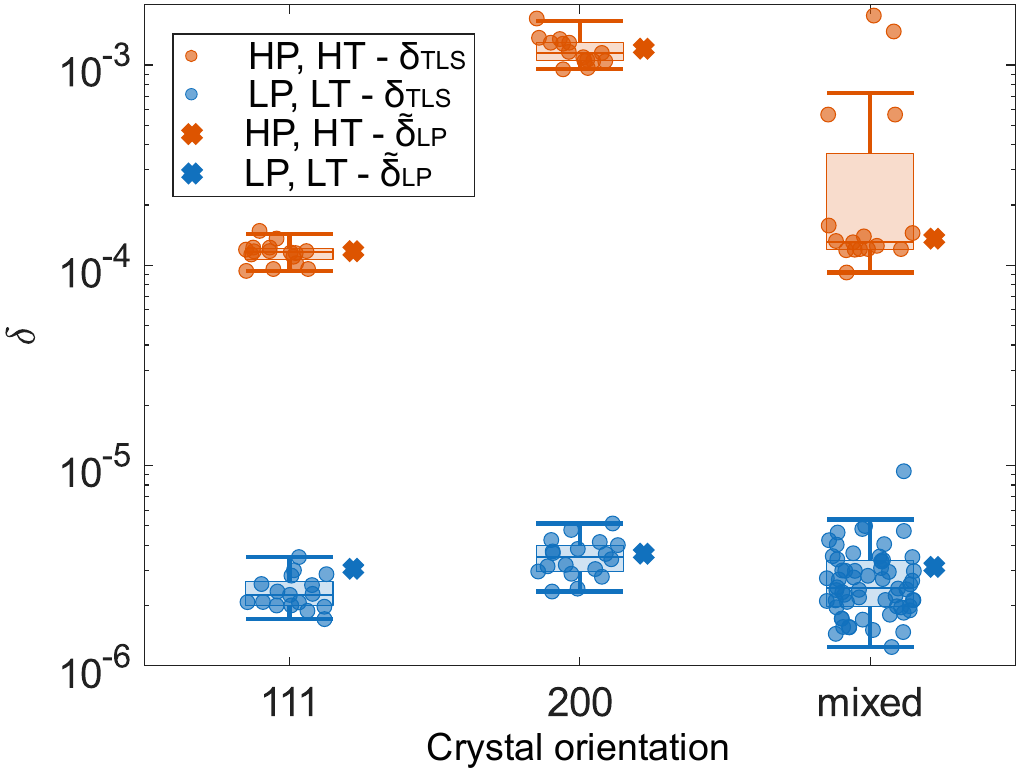}
    \caption{\label{fig:delta_TiN}$\delta_{\mathrm{TLS}}$ and $\tilde{\delta}_{\mathrm{LP}}$ for resonators structured from different TiN crystal orientations using the worst (HP+HT, red) and the best (LP+LT, blue) performing structuring methods tested.}
\end{figure}

The first part of the cryogenic characterization focuses on analyzing the influence of the TiN material on the RRR value, \( T_c \), and the quality factor of the resonators. Figure~\ref{fig:TcRRR} summarizes the first two of these three parameters.
All TiN thin films studied here exhibit a residual-resistance ratio (RRR) $>\,1$, characteristic of metallic behaviour, and the proximity of this value to unity indicates that the resistivity is dominated by defect scattering. The TiN(111) film and the TiN(200) film have similar RRR values. In contrast, the mixed‑orientation TiN film presents a slightly reduced RRR, despite having an average grain size larger than that of TiN(200). This observation rules out grain‑boundary density as the explanation for the difference in RRR.
Instead, the type of grain boundaries might be decisive. TiN(111) and TiN(200) films contain only homo‑oriented ((111)/(111) or (200)/(200) boundaries), whereas the mixed film (also) possesses hetero‑oriented (111)/(200) grain boundaries. The additional disorder introduced by these hetero‑interfaces might increase electron scattering and therefore lowers the RRR.
%Despite having a similar RRR TiN(111) exhibits a substantially higher room‑temperature resistivity than TiN(200) and TiNmixed.
The $T_c$ values of all the TiN films are in the region between $4.2\,$K and $5.2\,$K and therefore are only slightly lower than the bulk literature value of $5.8\,$K \cite{Rumble_CRC_2022}. 
\newline
The TLS and LP losses of the three TiN films are demonstrated in Figure~\ref{fig:delta_TiN}. For improved readability, only the median $\tilde{\delta}_{\mathrm{LP}}$ of the LP losses are shown.
The box plot visualizes data from single resonators as circles and the box and the bars visualize the interquartile range (IQR) with outliers above 1.5*IQR. 
We compare the HP/HT and LP/LT process combinations only, chosen to represent the extremes: the LP etch with LT strip, which yields the best performance with $\tilde{\delta}_{\mathrm{LP}}$ and $\delta_{\rm TLS}$ approaching $10^{-6}$, and the HP etch with HP strip, which yields the worst performance with both losses two to three orders of magnitude higher (details in following section). 
\newline
The TLS loss characteristics of the three TiN film orientations exhibit distinct trends. The TiN200 films generally exhibit higher losses compared to the TiN111 films, particularly for the HP/HT process combinations. The TiNmixed films demonstrate a median TLS loss $\tilde{\delta}_{\rm TLS}$ comparable to that of the TiN111 films, but with data points spanning (more) than the entire range observed for both TiN111 and TiN200 films. This trend is consistent across both structuring methods but becomes very evident when looking at the losses of the HP,HT resonators. 
The crystal orientation of the film cannot explain this behavior as the TiNmixed film preferably contains (200) oriented grains but mostly performs like the TiN(111) film.
One explanation for this behavior is the grain size. 
A smaller grain size results in a higher density of grain boundaries, which are susceptible to oxidation \cite{biznarova_mitigation_2024}. Since oxides are known to host TLS defects \cite{muller_towards_2019, oconnell_microwave_2008, altoe_localization_2022, chang_eliminating_2025, verjauw_investigation_2021, woods_determining_2019,  gupta_high_2026}, this is expected to contribute to the increased losses especially observed in the fine grained TiN200 resonators.
However, even with this explanation the question why the TiNmixed film mostly behaves like the TiN111 film remains unanswered.  
One possible explanation are exceptionally lossy homo‑oriented (200)/(200) grain boundaries in the TiN200 film.
TiN111 films, of course, do not contain such boundaries because they possess only homo‑oriented (111)/(111) grain boundaries.  
In contrast, TiNmixed films contain fewer (200)/(200) grain boundaries, since hetero‑oriented (200)/(111) boundaries must also be present.  
Consequently, only resonators that happen to incorporate (a large number) of (200)/(200) grain boundaries exhibit performance comparable to that of the TiN200 resonators.

%However, XRD measurements indicate that the grain sizes in the TiNmixed films are larger than those in the TiN200 films and similar to those of the TiN111 films. This observation likely explains the generally superior performance of the TiNmixed films and supports the hypothesis that grain boundaries and their associated oxides substantially contribute to TLS losses. 
%Consequently, TiNmixed resonators that incorporate more grain boundaries by chance exhibit performance comparable to that of the TiN200 resonators.
%Apart from the grain size alone (independent of the orientation of the grains) there is another  explanation for the worse performance of our TiN200 films.

\subsubsection{Influence of structuring processes}
\begin{figure}
    \includegraphics[width=\linewidth]{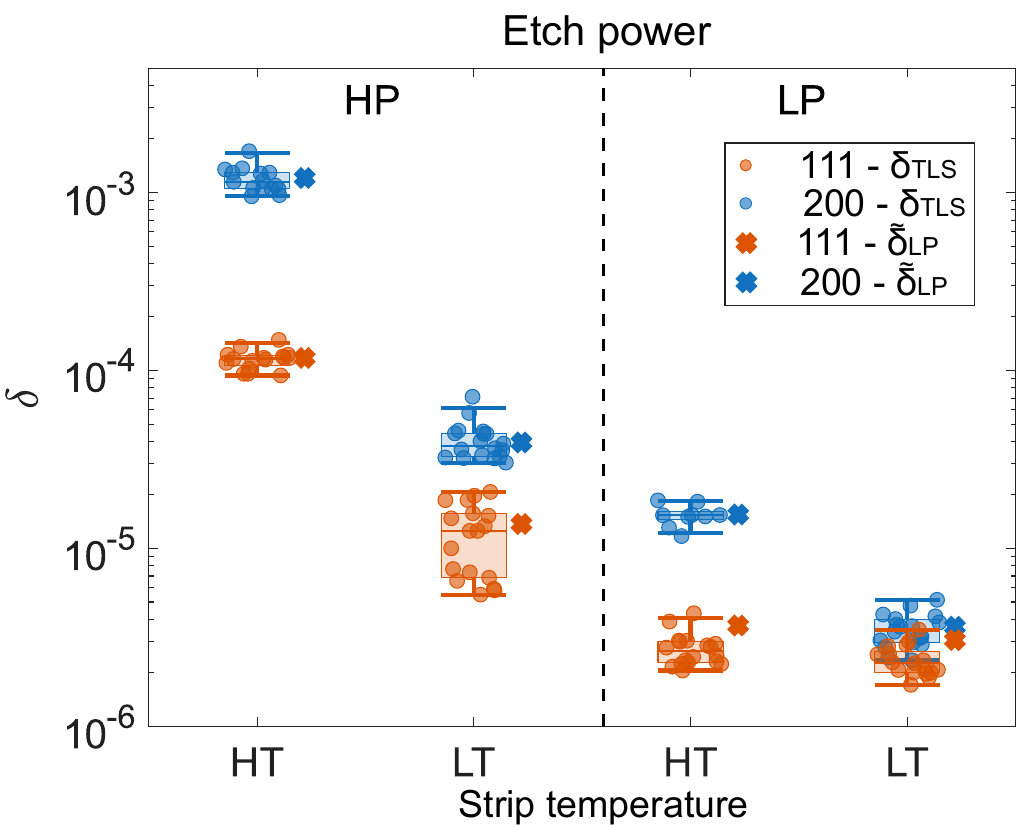}
    \caption{\label{fig:delta_etch}$\delta_{\mathrm{TLS}}$ and $\tilde{\delta}_{\mathrm{LP}}$ for TiN111 (red) and TiN200 (blue) resonators with different etch recipes (top) and strip temperatures (bottom).}
\end{figure}

To examine the influence of the structuring processes on the TLS and LP loss of the respective resonators Figure~\ref{fig:delta_etch} summarizes $\delta_{\mathrm{TLS}}$ for all four process combinations derived from pairing LP and HP etching with LT and HT resist stripping. The median for both, TLS and LP losses, are comparable for the respective process parameters.
As TiNmixed films are expected to exhibit behavior consistent with the trends described earlier, they are excluded from the discussion in this section. 
%The generally better performance of the TiN111 films compared to the TiN200 films, mentioned above, is again confirmed in this graph.
The dielectric substrate loss and the MS interface losses remain constant for the processing variations discussed here.
%since the Si substrates used are well-defined high-ohmic wafers, the dielectric substrate loss $\delta_{\mathrm{Si}}$ remains constant across all process variations under discussion. Additionally the TLS loss at the MS interface also remains constant within each deposition process. 
It should also be noted that for small overetch depths in the range of 0–100\,nm, the participation ratios of our designs decrease slightly with increasing overetch depth, as shown in Table~\ref{tab:PPR} because the electric field moves further away from lossy regions. 
\paragraph{Dry etching}
According to the point just made the HP etch, which introduces overetching into the silicon, should, if anything, result in lower losses. However, it is a clear result of this study that the HP etch generally introduces more losses into the system than the LP etch. This is the case for all four tested parameter variations.
This etch process influences two interfaces: the SA interface and the MA interface, but the MA interface only in the small region of the sidewall. %Distinguishing between these two effects is not possible with the current setup. 
The SA interface of the HP etching is rougher than that of the LP etching, as confirmed by the AFM measurements above. Additionally, the higher bias voltage in the HP etching might lead to stronger amorphization of the Si surface due to bombardment by accelerated ions. Rougher surfaces, as well as amorphous silicon, are prone to increased oxidation, and it is known that amorphous silicon exhibits higher dielectric loss than crystalline silicon \cite{quintana_characterization_2014, dunsworth_characterization_2017, murray_material_2021}.
To check whether the process conditions for the HP etch result in the implantation of boron into the Si surface \cite{sandberg_etch_2012}, a vapor phase decomposition (VPD) analysis of the Si surface was performed. However, the surface boron concentrations found for the HP process ($4.1 \cdot 10^{10}$\,cm$^{-2}$) and the LP process ($10.6 \cdot 10^{10}$\,cm$^{-2}$) are lower compared to the values for untreated silicon ($50.3 \cdot 10^{10}$\,cm$^{-2}$).
In addition to the increase in Si surface roughness, there is a substantial change in the sidewall profile (Fig.\,\ref{fig:TiNdepoSEM}), suggesting the poorer HP etch performance is also linked to sidewall alterations. The by far rougher LP-etched sidewall was previously connected to different chemical etch rates for grain boundaries and TiN grains. This implies that the LP etch tends to stop at existing grain boundaries, while the HP etch exposes "bulk" TiN which might develop a lossier oxide compared to grain boundaries afterwards. Alternatively, the HP etch forms polymer residues on the sidewalls, which are not fully removed by subsequent treatments and therefore increase the loss.
\paragraph{Resist stripping}
Figure~\ref{fig:delta_etch} clearly shows that the LT strip process results in lower losses compared to the HT strip process, consistently across all four process combinations. This processing step, being an H$_2$O vapor plasma step, is associated with the growth of thicker oxide layers at the SA and MA interfaces, which explains the worse performance of the respective resonators.
\newline
All in all, structuring processes that involve high‑temperature oxygen‑plasma treatment or exposure of non‑oxidized bulk material promote additional oxide growth at interfaces in general and at grain boundaries in particular, which increases the resonator losses.
This effect is especially pronounced for resonators fabricated from TiN200 films because this material is prone to develop lossy oxides.  This hypothesis (formulated in the previous chapter) is further supported by the observation that the loss difference between the HT and LT strip processes for TiN200 is much larger than that for TiN111.
The dominant contribution to the total loss can therefore be attributed to the MA‑interface region, while the Si interface (common to all four process combinations examined) shows negligible variation.
Moreover, amorphization, the presence of etch polymers on etched surfaces, and surface roughness are also suspected to increase the losses.
%Another explanatory approach for the poor performance of the TiN200 films might be that the deposition process generates a thicker MS region, which contributes more to the losses. However, this explanation can be ruled out, as the TLS and LP losses of the exact same film are minimal when the LP etch is combined with the LP strip process, and neither of these processes alters the MS region.

\subsubsection{Influence of BOE wet etching and time coupling}
\begin{figure}
    \includegraphics[width=\linewidth]{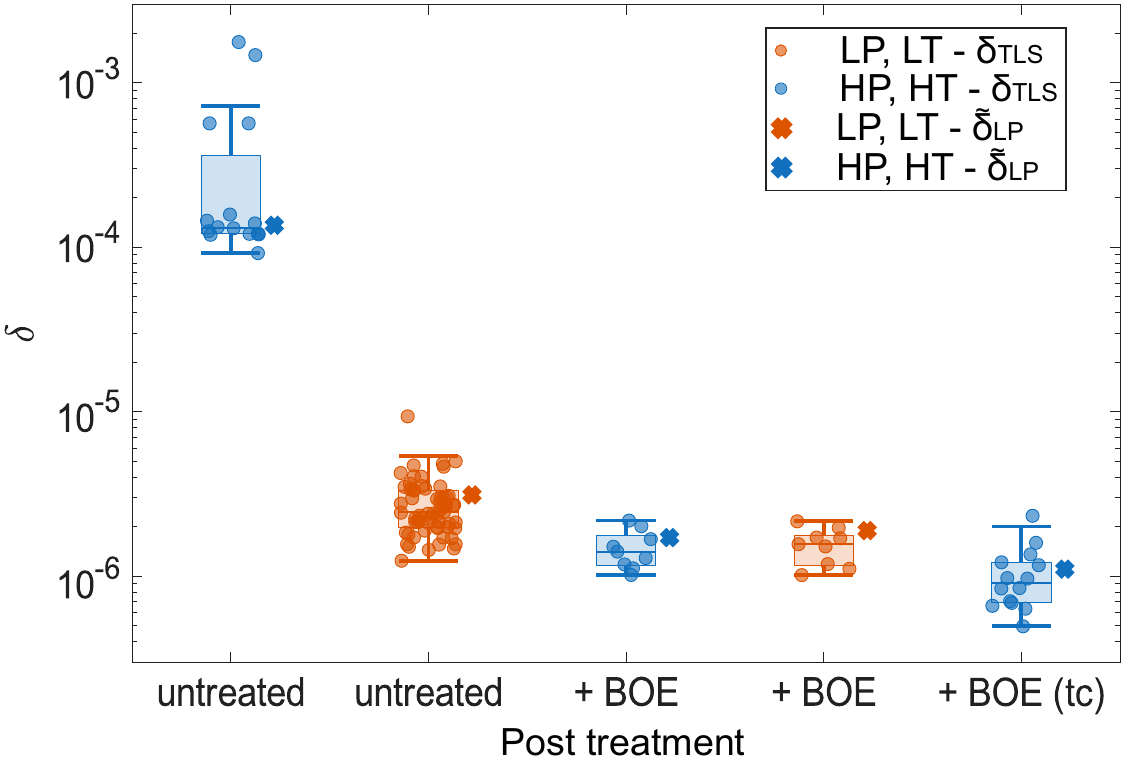}
\caption{\label{fig:delta_BOE}Influence of BOE treatment on $\delta_\mathrm{TLS}$ and $\tilde{\delta}_{\mathrm{LP}}$ of HP+HT (blue) and LP+LT (red) manufactured TiNmixed samples and in combination with 2 hour timecoupling (tc) between measurements and cooldown. }
\end{figure}

Since the TiNmixed films tend to span the range of quality factors expected for TiN111 and TiN200 films, this experiment was conducted exclusively with TiNmixed films. The process combinations used were the best-performing combination (LP etch and LT strip) and the worst-performing combination (HP etch and HT strip), as in previous experiments.
BOE etches native oxides on Si and TiN, affecting both SA and MA interfaces, including sidewalls and top surfaces.
The results of the experiment, where resonators made from TiNmixed films were BOE-etched and then the same films were remeasured, are shown in Figure~\ref{fig:delta_BOE}.
Both samples were cooled down within one day after etching, while an additional sample, etched with HP and stripped with HT, was cooled down approximately two hours after BOE etching (tc). 
Although the two structuring procedures under investigation initially exhibit $\delta_\mathrm{TLS}$ and $\tilde{\delta}_\mathrm{LP}$ values that differ by approximately two orders of magnitude, they converge to the same loss level after BOE etching. Time coupling further reduced $\tilde{\delta}_\mathrm{TLS}$ to $9.67\cdot10^{-7}$ and $\tilde{\delta}_\mathrm{LP}$ to $11.04\cdot10^{-7}$.
\\
This result aligns with the suggestion that the losses introduced by etching and stripping can be attributed to lossy oxides at the SA and MA interfaces, including oxidation at grain boundaries. The small difference observed between the 2-hour and 24-hour delays after BOE treatment indicates that reoxidation at room temperature saturates quickly. 
%The larger spread of the losses for the time coupled measurements however support the idea of grain (boundaries) with different crystal orientations show different oxidation behavior. 
Interestingly, the oxide resulting from the LT strip has only slightly more loss compared to the BOE treated sample, while being exposed to water vapor plasma.

\section{Conclusion}

A comprehensive experimental study was carried out to assess the dielectric losses of TiN resonators fabricated with three deposition processes, two etch processes, and two resist‑strip processes.  
We obtained two distinct polycrystalline TiN films with exclusive in‑plane orientations (111) and (200), as well as a mixed film that combines both orientations; all textures were verified by X‑ray diffraction (XRD).  
Resonators patterned from the three TiN films showed comparable minimal TLS and LP losses, with TiN111 delivering the lowest median loss and exhibiting noticeably higher resistance to re‑oxidation than TiN200 and TiNmixed.  
The chosen etch and resist‑strip processes exerted a far greater influence on the measured $\delta_{\mathrm{TLS}}$ and $\delta_{\mathrm{LP}}$ than the crystal orientation.  This demonstrates that, although the MS interface possesses the largest participation ratio, the lossy surface oxides at the SA and MA interfaces dominate the loss channels.  
For the LP/LT resonators, structuring‑induced losses originating from the SA and MA regions are substantially reduced, suggesting that the dominant loss mechanism may now shift toward the MS interface.  
By keeping the participation ratios essentially unchanged and comparing the various process combinations through structural analysis and cryogenic characterization, we identified the limiting losses as those introduced by steps that promote oxide growth on the MA and SA interfaces.  This hypothesis was confirmed with a BOE post‑treatment that removed the oxides formed during these steps, rendering the prior etch and resist‑strip choices irrelevant.  
Consequently, the BOE post‑treatment emerges as an effective method to eliminate TLS losses caused by the structuring process.  Ensuring that measured losses are not dominated by the fabrication method is essential for accurate characterization, especially when the goal is to isolate the impact of the superconducting layer properties on resonator dielectric losses.  
Combining the BOE dip with time‑coupled measurements yielded the lowest median TLS loss, $\tilde{\delta}_{\mathrm{TLS}} = 9.67\cdot10^{-7}$, and the lowest LP loss, $\tilde{\delta}_{\mathrm{LP}} = 11.04\cdot10^{-7}$, approaching an internal quality factor of one million, in line with values reported in the literature \cite{vissers_low_2010}.  
Overall, this broad study provides insight into the critical steps of an industry‑grade, 200‑mm CMOS‑compatible resonator workflow used for qubit fabrication.  

\begin{acknowledgements}
The authors gratefully acknowledge the support of TUM CRC for assistance with XRD measurements and extend their thanks to Felix Rucker, Sabrina Kressierer, and Oscar Gargiulo from kiutra GmbH for insightful discussions regarding cryogenic measurements.
Special thanks are due to Dr. Lars Nebrich for his valuable support with layout design. The authors also express their gratitude to M. Hahn and M. König for their contributions and constructive discussions during process development. Finally, the authors thank the entire Fraunhofer EMFT cleanroom staff for their expertise and dedication in ensuring high-quality fabrication. Cover image by Bernd Müller.

This work was funded by the Munich Quantum Valley (MQV) – Consortium Scalable Hardware and Systems Engineering (SHARE), funded by the Bavarian State Government with funds from the Hightech Agenda Bavaria, the Munich Quantum Valley Quantum Computer Demonstrator - Superconducting Qubits (MUNIQC-SC) 13N16188, funded by the Federal Ministry of Education and Research, Germany, and the Open Superconducting Quantum Computers (OpenSuperQPlus) Project - European Quantum Technology Flagship.
\end{acknowledgements}

\section*{Data Availability Statement}
The data that support the findings of
this study are available from the
corresponding author upon reasonable
request.

\bibliographystyle{aipnum4-1} % AIP numerical style
\bibliography{paper.bib}  % Your .bib file

\begin{thebibliography}{39}%
\makeatletter
\providecommand \@ifxundefined [1]{%
 \@ifx{#1\undefined}
}%
\providecommand \@ifnum [1]{%
 \ifnum #1\expandafter \@firstoftwo
 \else \expandafter \@secondoftwo
 \fi
}%
\providecommand \@ifx [1]{%
 \ifx #1\expandafter \@firstoftwo
 \else \expandafter \@secondoftwo
 \fi
}%
\providecommand \natexlab [1]{#1}%
\providecommand \enquote  [1]{``#1''}%
\providecommand \bibnamefont  [1]{#1}%
\providecommand \bibfnamefont [1]{#1}%
\providecommand \citenamefont [1]{#1}%
\providecommand \href@noop [0]{\@secondoftwo}%
\providecommand \href [0]{\begingroup \@sanitize@url \@href}%
\providecommand \@href[1]{\@@startlink{#1}\@@href}%
\providecommand \@@href[1]{\endgroup#1\@@endlink}%
\providecommand \@sanitize@url [0]{\catcode `\\12\catcode `\$12\catcode
  `\&12\catcode `\#12\catcode `\^12\catcode `\_12\catcode `\%12\relax}%
\providecommand \@@startlink[1]{}%
\providecommand \@@endlink[0]{}%
\providecommand \url  [0]{\begingroup\@sanitize@url \@url }%
\providecommand \@url [1]{\endgroup\@href {#1}{\urlprefix }}%
\providecommand \urlprefix  [0]{URL }%
\providecommand \Eprint [0]{\href }%
\providecommand \doibase [0]{http://dx.doi.org/}%
\providecommand \selectlanguage [0]{\@gobble}%
\providecommand \bibinfo  [0]{\@secondoftwo}%
\providecommand \bibfield  [0]{\@secondoftwo}%
\providecommand \translation [1]{[#1]}%
\providecommand \BibitemOpen [0]{}%
\providecommand \bibitemStop [0]{}%
\providecommand \bibitemNoStop [0]{.\EOS\space}%
\providecommand \EOS [0]{\spacefactor3000\relax}%
\providecommand \BibitemShut  [1]{\csname bibitem#1\endcsname}%
\let\auto@bib@innerbib\@empty
%</preamble>
\bibitem [{\citenamefont {Fowler}\ \emph {et~al.}(2012)\citenamefont {Fowler},
  \citenamefont {Mariantoni}, \citenamefont {Martinis},\ and\ \citenamefont
  {Cleland}}]{fowler_Surface_2012}%
  \BibitemOpen
  \bibfield  {author} {\bibinfo {author} {\bibfnamefont {A.~G.}\ \bibnamefont
  {Fowler}}, \bibinfo {author} {\bibfnamefont {M.}~\bibnamefont {Mariantoni}},
  \bibinfo {author} {\bibfnamefont {J.~M.}\ \bibnamefont {Martinis}}, \ and\
  \bibinfo {author} {\bibfnamefont {A.~N.}\ \bibnamefont {Cleland}},\ }\href
  {\doibase 10.1103/PhysRevA.86.032324} {\bibfield  {journal} {\bibinfo
  {journal} {Phys. Rev. A}\ }\textbf {\bibinfo {volume} {86}},\ \bibinfo
  {pages} {032324} (\bibinfo {year} {2012})}\BibitemShut {NoStop}%
\bibitem [{\citenamefont {Grover}(1997)}]{Grover_Quantum_1997}%
  \BibitemOpen
  \bibfield  {author} {\bibinfo {author} {\bibfnamefont {L.~K.}\ \bibnamefont
  {Grover}},\ }\href {\doibase 10.1103/PhysRevLett.79.325} {\bibfield
  {journal} {\bibinfo  {journal} {Phys. Rev. Lett.}\ }\textbf {\bibinfo
  {volume} {79}},\ \bibinfo {pages} {325} (\bibinfo {year} {1997})}\BibitemShut
  {NoStop}%
\bibitem [{\citenamefont {Yost}\ \emph {et~al.}(2020)\citenamefont {Yost},
  \citenamefont {Schwartz}, \citenamefont {Mallek}, \citenamefont {Rosenberg},
  \citenamefont {Stull},\ and\ \citenamefont {et~al.}}]{Yost_Solid_2020}%
  \BibitemOpen
  \bibfield  {author} {\bibinfo {author} {\bibfnamefont {D.~R.~W.}\
  \bibnamefont {Yost}}, \bibinfo {author} {\bibfnamefont {M.~E.}\ \bibnamefont
  {Schwartz}}, \bibinfo {author} {\bibfnamefont {J.}~\bibnamefont {Mallek}},
  \bibinfo {author} {\bibfnamefont {D.}~\bibnamefont {Rosenberg}}, \bibinfo
  {author} {\bibfnamefont {C.}~\bibnamefont {Stull}}, \ and\ \bibinfo {author}
  {\bibnamefont {et~al.}},\ }\href {\doibase 10.1038/s41534-020-00289-8}
  {\bibfield  {journal} {\bibinfo  {journal} {npj Quantum Information}\
  }\textbf {\bibinfo {volume} {6}},\ \bibinfo {pages} {59} (\bibinfo {year}
  {2020})}\BibitemShut {NoStop}%
\bibitem [{\citenamefont {Arute}\ \emph {et~al.}(2019)\citenamefont {Arute},
  \citenamefont {Arya}, \citenamefont {Babbush}, \citenamefont {Bacon},
  \citenamefont {Bardin},\ and\ \citenamefont {et~al.}}]{Arute_Quantum_2019}%
  \BibitemOpen
  \bibfield  {author} {\bibinfo {author} {\bibfnamefont {F.}~\bibnamefont
  {Arute}}, \bibinfo {author} {\bibfnamefont {K.}~\bibnamefont {Arya}},
  \bibinfo {author} {\bibfnamefont {R.}~\bibnamefont {Babbush}}, \bibinfo
  {author} {\bibfnamefont {D.}~\bibnamefont {Bacon}}, \bibinfo {author}
  {\bibfnamefont {J.~C.}\ \bibnamefont {Bardin}}, \ and\ \bibinfo {author}
  {\bibnamefont {et~al.}},\ }\href {\doibase 10.1038/s41586-019-1666-5}
  {\bibfield  {journal} {\bibinfo  {journal} {Nature}\ }\textbf {\bibinfo
  {volume} {574}},\ \bibinfo {pages} {505} (\bibinfo {year}
  {2019})}\BibitemShut {NoStop}%
\bibitem [{\citenamefont {Gao}\ \emph {et~al.}(2008)\citenamefont {Gao},
  \citenamefont {Daal}, \citenamefont {Vayonakis}, \citenamefont {Kumar},
  \citenamefont {Zmuidzinas}, \citenamefont {Sadoulet}, \citenamefont {Mazin},
  \citenamefont {Day},\ and\ \citenamefont {Leduc}}]{gao_experimental_2008}%
  \BibitemOpen
  \bibfield  {author} {\bibinfo {author} {\bibfnamefont {J.}~\bibnamefont
  {Gao}}, \bibinfo {author} {\bibfnamefont {M.}~\bibnamefont {Daal}}, \bibinfo
  {author} {\bibfnamefont {A.}~\bibnamefont {Vayonakis}}, \bibinfo {author}
  {\bibfnamefont {S.}~\bibnamefont {Kumar}}, \bibinfo {author} {\bibfnamefont
  {J.}~\bibnamefont {Zmuidzinas}}, \bibinfo {author} {\bibfnamefont
  {B.}~\bibnamefont {Sadoulet}}, \bibinfo {author} {\bibfnamefont {B.~A.}\
  \bibnamefont {Mazin}}, \bibinfo {author} {\bibfnamefont {P.~K.}\ \bibnamefont
  {Day}}, \ and\ \bibinfo {author} {\bibfnamefont {H.~G.}\ \bibnamefont
  {Leduc}},\ }\href {\doibase 10.1063/1.2906373} {\bibfield  {journal}
  {\bibinfo  {journal} {Applied Physics Letters}\ }\textbf {\bibinfo {volume}
  {92}},\ \bibinfo {pages} {152505} (\bibinfo {year} {2008})}\BibitemShut
  {NoStop}%
\bibitem [{\citenamefont {Schoof}\ \emph {et~al.}(2024)\citenamefont {Schoof},
  \citenamefont {Singer}, \citenamefont {Lang}, \citenamefont {Gupta},
  \citenamefont {Zahn}, \citenamefont {Weber},\ and\ \citenamefont
  {Tornow}}]{schoof_development_2024_2}%
  \BibitemOpen
  \bibfield  {author} {\bibinfo {author} {\bibfnamefont {B.}~\bibnamefont
  {Schoof}}, \bibinfo {author} {\bibfnamefont {M.}~\bibnamefont {Singer}},
  \bibinfo {author} {\bibfnamefont {S.}~\bibnamefont {Lang}}, \bibinfo {author}
  {\bibfnamefont {H.}~\bibnamefont {Gupta}}, \bibinfo {author} {\bibfnamefont
  {D.}~\bibnamefont {Zahn}}, \bibinfo {author} {\bibfnamefont {J.}~\bibnamefont
  {Weber}}, \ and\ \bibinfo {author} {\bibfnamefont {M.}~\bibnamefont
  {Tornow}},\ }in\ \href {\doibase 10.1109/QCE60285.2024.00145} {\emph
  {\bibinfo {booktitle} {2024 IEEE International Conference on Quantum
  Computing and Engineering (QCE)}}}\ (\bibinfo  {publisher} {IEEE Computer
  Society},\ \bibinfo {address} {Los Alamitos, CA, USA},\ \bibinfo {year}
  {2024})\ pp.\ \bibinfo {pages} {1228--1232}\BibitemShut {NoStop}%
\bibitem [{\citenamefont {Sandberg}\ \emph {et~al.}(2012)\citenamefont
  {Sandberg}, \citenamefont {Vissers}, \citenamefont {Kline}, \citenamefont
  {Weides}, \citenamefont {Gao}, \citenamefont {Wisbey},\ and\ \citenamefont
  {Pappas}}]{sandberg_etch_2012}%
  \BibitemOpen
  \bibfield  {author} {\bibinfo {author} {\bibfnamefont {M.}~\bibnamefont
  {Sandberg}}, \bibinfo {author} {\bibfnamefont {M.~R.}\ \bibnamefont
  {Vissers}}, \bibinfo {author} {\bibfnamefont {J.~S.}\ \bibnamefont {Kline}},
  \bibinfo {author} {\bibfnamefont {M.}~\bibnamefont {Weides}}, \bibinfo
  {author} {\bibfnamefont {J.}~\bibnamefont {Gao}}, \bibinfo {author}
  {\bibfnamefont {D.~S.}\ \bibnamefont {Wisbey}}, \ and\ \bibinfo {author}
  {\bibfnamefont {D.~P.}\ \bibnamefont {Pappas}},\ }\href {\doibase
  10.1063/1.4729623} {\bibfield  {journal} {\bibinfo  {journal} {Applied
  Physics Letters}\ }\textbf {\bibinfo {volume} {100}},\ \bibinfo {pages}
  {262605} (\bibinfo {year} {2012})}\BibitemShut {NoStop}%
\bibitem [{\citenamefont {Vissers}\ \emph {et~al.}(2010)\citenamefont
  {Vissers}, \citenamefont {Gao}, \citenamefont {Wisbey}, \citenamefont {Hite},
  \citenamefont {Tsuei}, \citenamefont {Corcoles}, \citenamefont {Steffen},\
  and\ \citenamefont {Pappas}}]{vissers_low_2010}%
  \BibitemOpen
  \bibfield  {author} {\bibinfo {author} {\bibfnamefont {M.~R.}\ \bibnamefont
  {Vissers}}, \bibinfo {author} {\bibfnamefont {J.}~\bibnamefont {Gao}},
  \bibinfo {author} {\bibfnamefont {D.~S.}\ \bibnamefont {Wisbey}}, \bibinfo
  {author} {\bibfnamefont {D.~A.}\ \bibnamefont {Hite}}, \bibinfo {author}
  {\bibfnamefont {C.~C.}\ \bibnamefont {Tsuei}}, \bibinfo {author}
  {\bibfnamefont {A.~D.}\ \bibnamefont {Corcoles}}, \bibinfo {author}
  {\bibfnamefont {M.}~\bibnamefont {Steffen}}, \ and\ \bibinfo {author}
  {\bibfnamefont {D.~P.}\ \bibnamefont {Pappas}},\ }\href {\doibase
  10.1063/1.3517252} {\bibfield  {journal} {\bibinfo  {journal} {Applied
  Physics Letters}\ }\textbf {\bibinfo {volume} {97}},\ \bibinfo {pages}
  {232509} (\bibinfo {year} {2010})}\BibitemShut {NoStop}%
\bibitem [{\citenamefont {Ohya}\ \emph {et~al.}(2014)\citenamefont {Ohya},
  \citenamefont {Chiaro}, \citenamefont {Megrant}, \citenamefont {Neill},
  \citenamefont {Barends}, \citenamefont {Chen}, \citenamefont {Kelly},
  \citenamefont {Low}, \citenamefont {Mutus}, \citenamefont
  {O{\textquoteright}Malley}, \citenamefont {Roushan}, \citenamefont {Sank},
  \citenamefont {Vainsencher}, \citenamefont {Wenner}, \citenamefont {White},
  \citenamefont {Yin}, \citenamefont {Schultz}, \citenamefont {Palmstr{\o}m},
  \citenamefont {Mazin}, \citenamefont {Cleland},\ and\ \citenamefont
  {Martinis}}]{ohya_room_2014}%
  \BibitemOpen
  \bibfield  {author} {\bibinfo {author} {\bibfnamefont {S.}~\bibnamefont
  {Ohya}}, \bibinfo {author} {\bibfnamefont {B.}~\bibnamefont {Chiaro}},
  \bibinfo {author} {\bibfnamefont {A.}~\bibnamefont {Megrant}}, \bibinfo
  {author} {\bibfnamefont {C.}~\bibnamefont {Neill}}, \bibinfo {author}
  {\bibfnamefont {R.}~\bibnamefont {Barends}}, \bibinfo {author} {\bibfnamefont
  {Y.}~\bibnamefont {Chen}}, \bibinfo {author} {\bibfnamefont {J.}~\bibnamefont
  {Kelly}}, \bibinfo {author} {\bibfnamefont {D.}~\bibnamefont {Low}}, \bibinfo
  {author} {\bibfnamefont {J.}~\bibnamefont {Mutus}}, \bibinfo {author}
  {\bibfnamefont {P.~J.~J.}\ \bibnamefont {O{\textquoteright}Malley}}, \bibinfo
  {author} {\bibfnamefont {P.}~\bibnamefont {Roushan}}, \bibinfo {author}
  {\bibfnamefont {D.}~\bibnamefont {Sank}}, \bibinfo {author} {\bibfnamefont
  {A.}~\bibnamefont {Vainsencher}}, \bibinfo {author} {\bibfnamefont
  {J.}~\bibnamefont {Wenner}}, \bibinfo {author} {\bibfnamefont {T.~C.}\
  \bibnamefont {White}}, \bibinfo {author} {\bibfnamefont {Y.}~\bibnamefont
  {Yin}}, \bibinfo {author} {\bibfnamefont {B.~D.}\ \bibnamefont {Schultz}},
  \bibinfo {author} {\bibfnamefont {C.~J.}\ \bibnamefont {Palmstr{\o}m}},
  \bibinfo {author} {\bibfnamefont {B.~A.}\ \bibnamefont {Mazin}}, \bibinfo
  {author} {\bibfnamefont {A.~N.}\ \bibnamefont {Cleland}}, \ and\ \bibinfo
  {author} {\bibfnamefont {J.~M.}\ \bibnamefont {Martinis}},\ }\href {\doibase
  10.1088/0953-2048/27/1/015009} {\bibfield  {journal} {\bibinfo  {journal}
  {Supercond. Sci. Technol.}\ }\textbf {\bibinfo {volume} {27}},\ \bibinfo
  {pages} {015009} (\bibinfo {year} {2014})}\BibitemShut {NoStop}%
\bibitem [{\citenamefont {Chang}\ \emph {et~al.}(2013)\citenamefont {Chang},
  \citenamefont {Vissers}, \citenamefont {C{\'o}rcoles}, \citenamefont
  {Sandberg}, \citenamefont {Gao}, \citenamefont {Abraham}, \citenamefont
  {Chow}, \citenamefont {Gambetta}, \citenamefont {Beth~Rothwell},
  \citenamefont {Keefe}, \citenamefont {Steffen},\ and\ \citenamefont
  {Pappas}}]{chang_improved_2013}%
  \BibitemOpen
  \bibfield  {author} {\bibinfo {author} {\bibfnamefont {J.~B.}\ \bibnamefont
  {Chang}}, \bibinfo {author} {\bibfnamefont {M.~R.}\ \bibnamefont {Vissers}},
  \bibinfo {author} {\bibfnamefont {A.~D.}\ \bibnamefont {C{\'o}rcoles}},
  \bibinfo {author} {\bibfnamefont {M.}~\bibnamefont {Sandberg}}, \bibinfo
  {author} {\bibfnamefont {J.}~\bibnamefont {Gao}}, \bibinfo {author}
  {\bibfnamefont {D.~W.}\ \bibnamefont {Abraham}}, \bibinfo {author}
  {\bibfnamefont {J.~M.}\ \bibnamefont {Chow}}, \bibinfo {author}
  {\bibfnamefont {J.~M.}\ \bibnamefont {Gambetta}}, \bibinfo {author}
  {\bibfnamefont {M.}~\bibnamefont {Beth~Rothwell}}, \bibinfo {author}
  {\bibfnamefont {G.~A.}\ \bibnamefont {Keefe}}, \bibinfo {author}
  {\bibfnamefont {M.}~\bibnamefont {Steffen}}, \ and\ \bibinfo {author}
  {\bibfnamefont {D.~P.}\ \bibnamefont {Pappas}},\ }\href {\doibase
  10.1063/1.4813269} {\bibfield  {journal} {\bibinfo  {journal} {Applied
  Physics Letters}\ }\textbf {\bibinfo {volume} {103}},\ \bibinfo {pages}
  {012602} (\bibinfo {year} {2013})}\BibitemShut {NoStop}%
\bibitem [{\citenamefont {Calusine}\ \emph {et~al.}(2018)\citenamefont
  {Calusine}, \citenamefont {Melville}, \citenamefont {Woods}, \citenamefont
  {Das}, \citenamefont {Stull}, \citenamefont {Bolkhovsky}, \citenamefont
  {Braje}, \citenamefont {Hover}, \citenamefont {Kim}, \citenamefont {Miloshi},
  \citenamefont {Rosenberg}, \citenamefont {Sevi}, \citenamefont {Yoder},
  \citenamefont {Dauler},\ and\ \citenamefont
  {Oliver}}]{calusine_analysis_2018}%
  \BibitemOpen
  \bibfield  {author} {\bibinfo {author} {\bibfnamefont {G.}~\bibnamefont
  {Calusine}}, \bibinfo {author} {\bibfnamefont {A.}~\bibnamefont {Melville}},
  \bibinfo {author} {\bibfnamefont {W.}~\bibnamefont {Woods}}, \bibinfo
  {author} {\bibfnamefont {R.}~\bibnamefont {Das}}, \bibinfo {author}
  {\bibfnamefont {C.}~\bibnamefont {Stull}}, \bibinfo {author} {\bibfnamefont
  {V.}~\bibnamefont {Bolkhovsky}}, \bibinfo {author} {\bibfnamefont
  {D.}~\bibnamefont {Braje}}, \bibinfo {author} {\bibfnamefont
  {D.}~\bibnamefont {Hover}}, \bibinfo {author} {\bibfnamefont {D.~K.}\
  \bibnamefont {Kim}}, \bibinfo {author} {\bibfnamefont {X.}~\bibnamefont
  {Miloshi}}, \bibinfo {author} {\bibfnamefont {D.}~\bibnamefont {Rosenberg}},
  \bibinfo {author} {\bibfnamefont {A.}~\bibnamefont {Sevi}}, \bibinfo {author}
  {\bibfnamefont {J.~L.}\ \bibnamefont {Yoder}}, \bibinfo {author}
  {\bibfnamefont {E.}~\bibnamefont {Dauler}}, \ and\ \bibinfo {author}
  {\bibfnamefont {W.~D.}\ \bibnamefont {Oliver}},\ }\href {\doibase
  10.1063/1.5006888} {\bibfield  {journal} {\bibinfo  {journal} {Applied
  Physics Letters}\ }\textbf {\bibinfo {volume} {112}},\ \bibinfo {pages}
  {062601} (\bibinfo {year} {2018})}\BibitemShut {NoStop}%
\bibitem [{\citenamefont {Amin}\ \emph {et~al.}(2022)\citenamefont {Amin},
  \citenamefont {Ladner}, \citenamefont {Jourdan}, \citenamefont {Hentz},
  \citenamefont {Roch},\ and\ \citenamefont {Renard}}]{amin_loss_2022}%
  \BibitemOpen
  \bibfield  {author} {\bibinfo {author} {\bibfnamefont {K.~R.}\ \bibnamefont
  {Amin}}, \bibinfo {author} {\bibfnamefont {C.}~\bibnamefont {Ladner}},
  \bibinfo {author} {\bibfnamefont {G.}~\bibnamefont {Jourdan}}, \bibinfo
  {author} {\bibfnamefont {S.}~\bibnamefont {Hentz}}, \bibinfo {author}
  {\bibfnamefont {N.}~\bibnamefont {Roch}}, \ and\ \bibinfo {author}
  {\bibfnamefont {J.}~\bibnamefont {Renard}},\ }\href {\doibase
  10.1063/5.0086019} {\bibfield  {journal} {\bibinfo  {journal} {Applied
  Physics Letters}\ }\textbf {\bibinfo {volume} {120}},\ \bibinfo {pages}
  {164001} (\bibinfo {year} {2022})}\BibitemShut {NoStop}%
\bibitem [{\citenamefont {Martinis}(2009)}]{martinis_superconducting_2009}%
  \BibitemOpen
  \bibfield  {author} {\bibinfo {author} {\bibfnamefont {J.~M.}\ \bibnamefont
  {Martinis}},\ }\href {\doibase 10.1007/s11128-009-0105-1} {\bibfield
  {journal} {\bibinfo  {journal} {Quantum Inf Process}\ }\textbf {\bibinfo
  {volume} {8}},\ \bibinfo {pages} {81} (\bibinfo {year} {2009})}\BibitemShut
  {NoStop}%
\bibitem [{\citenamefont {Harrington}\ and\ \citenamefont
  {Santiso}(2021)}]{harrington_back_2021}%
  \BibitemOpen
  \bibfield  {author} {\bibinfo {author} {\bibfnamefont {G.~F.}\ \bibnamefont
  {Harrington}}\ and\ \bibinfo {author} {\bibfnamefont {J.}~\bibnamefont
  {Santiso}},\ }\href {\doibase 10.1007/s10832-021-00263-6} {\bibfield
  {journal} {\bibinfo  {journal} {J Electroceram}\ }\textbf {\bibinfo {volume}
  {47}},\ \bibinfo {pages} {141} (\bibinfo {year} {2021})}\BibitemShut
  {NoStop}%
\bibitem [{\citenamefont {Gross}\ and\ \citenamefont
  {Marx}(2018)}]{Gross_Festkörperphysik_2018}%
  \BibitemOpen
  \bibfield  {author} {\bibinfo {author} {\bibfnamefont {R.}~\bibnamefont
  {Gross}}\ and\ \bibinfo {author} {\bibfnamefont {A.}~\bibnamefont {Marx}},\
  }\href {\doibase doi:10.1515/9783110559187} {\emph {\bibinfo {title}
  {Festkörperphysik}}}\ (\bibinfo  {publisher} {De Gruyter},\ \bibinfo
  {address} {Berlin, Boston},\ \bibinfo {year} {2018})\BibitemShut {NoStop}%
\bibitem [{\citenamefont {Simons}(2001)}]{simons_coplanar_2001}%
  \BibitemOpen
  \bibfield  {author} {\bibinfo {author} {\bibfnamefont {R.~N.}\ \bibnamefont
  {Simons}},\ }\href {\doibase 10.1002/0471224758} {\emph {\bibinfo {title}
  {Coplanar {Waveguide} {Circuits}, {Components}, and {Systems}}}},\ \bibinfo
  {edition} {1st}\ ed.\ (\bibinfo  {publisher} {Wiley},\ \bibinfo {year}
  {2001})\BibitemShut {NoStop}%
\bibitem [{\citenamefont {G{\"o}ppl}\ \emph {et~al.}(2008)\citenamefont
  {G{\"o}ppl}, \citenamefont {Fragner}, \citenamefont {Baur}, \citenamefont
  {Bianchetti}, \citenamefont {Filipp}, \citenamefont {Fink}, \citenamefont
  {Leek}, \citenamefont {Puebla}, \citenamefont {Steffen},\ and\ \citenamefont
  {Wallraff}}]{goppl_coplanar_2008}%
  \BibitemOpen
  \bibfield  {author} {\bibinfo {author} {\bibfnamefont {M.}~\bibnamefont
  {G{\"o}ppl}}, \bibinfo {author} {\bibfnamefont {A.}~\bibnamefont {Fragner}},
  \bibinfo {author} {\bibfnamefont {M.}~\bibnamefont {Baur}}, \bibinfo {author}
  {\bibfnamefont {R.}~\bibnamefont {Bianchetti}}, \bibinfo {author}
  {\bibfnamefont {S.}~\bibnamefont {Filipp}}, \bibinfo {author} {\bibfnamefont
  {J.~M.}\ \bibnamefont {Fink}}, \bibinfo {author} {\bibfnamefont {P.~J.}\
  \bibnamefont {Leek}}, \bibinfo {author} {\bibfnamefont {G.}~\bibnamefont
  {Puebla}}, \bibinfo {author} {\bibfnamefont {L.}~\bibnamefont {Steffen}}, \
  and\ \bibinfo {author} {\bibfnamefont {A.}~\bibnamefont {Wallraff}},\ }\href
  {\doibase 10.1063/1.3010859} {\bibfield  {journal} {\bibinfo  {journal}
  {Journal of Applied Physics}\ }\textbf {\bibinfo {volume} {104}},\ \bibinfo
  {pages} {113904} (\bibinfo {year} {2008})}\BibitemShut {NoStop}%
\bibitem [{\citenamefont {Probst}\ \emph {et~al.}(2015)\citenamefont {Probst},
  \citenamefont {Song}, \citenamefont {Bushev}, \citenamefont {Ustinov},\ and\
  \citenamefont {Weides}}]{probst_efficient_2015}%
  \BibitemOpen
  \bibfield  {author} {\bibinfo {author} {\bibfnamefont {S.}~\bibnamefont
  {Probst}}, \bibinfo {author} {\bibfnamefont {F.~B.}\ \bibnamefont {Song}},
  \bibinfo {author} {\bibfnamefont {P.~A.}\ \bibnamefont {Bushev}}, \bibinfo
  {author} {\bibfnamefont {A.~V.}\ \bibnamefont {Ustinov}}, \ and\ \bibinfo
  {author} {\bibfnamefont {M.}~\bibnamefont {Weides}},\ }\href {\doibase
  10.1063/1.4907935} {\bibfield  {journal} {\bibinfo  {journal} {Review of
  Scientific Instruments}\ }\textbf {\bibinfo {volume} {86}},\ \bibinfo {pages}
  {024706} (\bibinfo {year} {2015})}\BibitemShut {NoStop}%
\bibitem [{\citenamefont {Bruno}\ \emph {et~al.}(2015)\citenamefont {Bruno},
  \citenamefont {de~Lange}, \citenamefont {Asaad}, \citenamefont {van~der
  Enden}, \citenamefont {Langford},\ and\ \citenamefont
  {DiCarlo}}]{bruno_surface_2015}%
  \BibitemOpen
  \bibfield  {author} {\bibinfo {author} {\bibfnamefont {A.}~\bibnamefont
  {Bruno}}, \bibinfo {author} {\bibfnamefont {G.}~\bibnamefont {de~Lange}},
  \bibinfo {author} {\bibfnamefont {S.}~\bibnamefont {Asaad}}, \bibinfo
  {author} {\bibfnamefont {K.~L.}\ \bibnamefont {van~der Enden}}, \bibinfo
  {author} {\bibfnamefont {N.~K.}\ \bibnamefont {Langford}}, \ and\ \bibinfo
  {author} {\bibfnamefont {L.}~\bibnamefont {DiCarlo}},\ }\href {\doibase
  10.1063/1.4919761} {\bibfield  {journal} {\bibinfo  {journal} {Applied
  Physics Letters}\ }\textbf {\bibinfo {volume} {106}},\ \bibinfo {pages}
  {182601} (\bibinfo {year} {2015})}\BibitemShut {NoStop}%
\bibitem [{\citenamefont {M{\"u}ller}, \citenamefont {Cole},\ and\
  \citenamefont {Lisenfeld}(2019)}]{muller_towards_2019}%
  \BibitemOpen
  \bibfield  {author} {\bibinfo {author} {\bibfnamefont {C.}~\bibnamefont
  {M{\"u}ller}}, \bibinfo {author} {\bibfnamefont {J.~H.}\ \bibnamefont
  {Cole}}, \ and\ \bibinfo {author} {\bibfnamefont {J.}~\bibnamefont
  {Lisenfeld}},\ }\href {\doibase 10.1088/1361-6633/ab3a7e} {\bibfield
  {journal} {\bibinfo  {journal} {Rep. Prog. Phys.}\ }\textbf {\bibinfo
  {volume} {82}},\ \bibinfo {pages} {124501} (\bibinfo {year}
  {2019})}\BibitemShut {NoStop}%
\bibitem [{\citenamefont {Martinis}\ \emph {et~al.}(2005)\citenamefont
  {Martinis}, \citenamefont {Cooper}, \citenamefont {McDermott}, \citenamefont
  {Steffen}, \citenamefont {Ansmann}, \citenamefont {Osborn}, \citenamefont
  {Cicak}, \citenamefont {Oh}, \citenamefont {Pappas}, \citenamefont
  {Simmonds},\ and\ \citenamefont {Yu}}]{martinis_decoherence_2005}%
  \BibitemOpen
  \bibfield  {author} {\bibinfo {author} {\bibfnamefont {J.~M.}\ \bibnamefont
  {Martinis}}, \bibinfo {author} {\bibfnamefont {K.~B.}\ \bibnamefont
  {Cooper}}, \bibinfo {author} {\bibfnamefont {R.}~\bibnamefont {McDermott}},
  \bibinfo {author} {\bibfnamefont {M.}~\bibnamefont {Steffen}}, \bibinfo
  {author} {\bibfnamefont {M.}~\bibnamefont {Ansmann}}, \bibinfo {author}
  {\bibfnamefont {K.~D.}\ \bibnamefont {Osborn}}, \bibinfo {author}
  {\bibfnamefont {K.}~\bibnamefont {Cicak}}, \bibinfo {author} {\bibfnamefont
  {S.}~\bibnamefont {Oh}}, \bibinfo {author} {\bibfnamefont {D.~P.}\
  \bibnamefont {Pappas}}, \bibinfo {author} {\bibfnamefont {R.~W.}\
  \bibnamefont {Simmonds}}, \ and\ \bibinfo {author} {\bibfnamefont {C.~C.}\
  \bibnamefont {Yu}},\ }\href {\doibase 10.1103/PhysRevLett.95.210503}
  {\bibfield  {journal} {\bibinfo  {journal} {Phys. Rev. Lett.}\ }\textbf
  {\bibinfo {volume} {95}},\ \bibinfo {pages} {210503} (\bibinfo {year}
  {2005})}\BibitemShut {NoStop}%
\bibitem [{\citenamefont {Wang}\ \emph {et~al.}(2015)\citenamefont {Wang},
  \citenamefont {Axline}, \citenamefont {Gao}, \citenamefont {Brecht},
  \citenamefont {Chu}, \citenamefont {Frunzio}, \citenamefont {Devoret},\ and\
  \citenamefont {Schoelkopf}}]{wang_surface_2015}%
  \BibitemOpen
  \bibfield  {author} {\bibinfo {author} {\bibfnamefont {C.}~\bibnamefont
  {Wang}}, \bibinfo {author} {\bibfnamefont {C.}~\bibnamefont {Axline}},
  \bibinfo {author} {\bibfnamefont {Y.~Y.}\ \bibnamefont {Gao}}, \bibinfo
  {author} {\bibfnamefont {T.}~\bibnamefont {Brecht}}, \bibinfo {author}
  {\bibfnamefont {Y.}~\bibnamefont {Chu}}, \bibinfo {author} {\bibfnamefont
  {L.}~\bibnamefont {Frunzio}}, \bibinfo {author} {\bibfnamefont {M.~H.}\
  \bibnamefont {Devoret}}, \ and\ \bibinfo {author} {\bibfnamefont {R.~J.}\
  \bibnamefont {Schoelkopf}},\ }\href {\doibase 10.1063/1.4934486} {\bibfield
  {journal} {\bibinfo  {journal} {Applied Physics Letters}\ }\textbf {\bibinfo
  {volume} {107}},\ \bibinfo {pages} {162601} (\bibinfo {year}
  {2015})}\BibitemShut {NoStop}%
\bibitem [{\citenamefont {Ganjam}\ \emph {et~al.}(2024)\citenamefont {Ganjam},
  \citenamefont {Wang}, \citenamefont {Lu}, \citenamefont {Banerjee},
  \citenamefont {Lei}, \citenamefont {Krayzman}, \citenamefont {Kisslinger},
  \citenamefont {Zhou}, \citenamefont {Li}, \citenamefont {Jia}, \citenamefont
  {Liu}, \citenamefont {Frunzio},\ and\ \citenamefont
  {Schoelkopf}}]{ganjam_surpassing_2024}%
  \BibitemOpen
  \bibfield  {author} {\bibinfo {author} {\bibfnamefont {S.}~\bibnamefont
  {Ganjam}}, \bibinfo {author} {\bibfnamefont {Y.}~\bibnamefont {Wang}},
  \bibinfo {author} {\bibfnamefont {Y.}~\bibnamefont {Lu}}, \bibinfo {author}
  {\bibfnamefont {A.}~\bibnamefont {Banerjee}}, \bibinfo {author}
  {\bibfnamefont {C.~U.}\ \bibnamefont {Lei}}, \bibinfo {author} {\bibfnamefont
  {L.}~\bibnamefont {Krayzman}}, \bibinfo {author} {\bibfnamefont
  {K.}~\bibnamefont {Kisslinger}}, \bibinfo {author} {\bibfnamefont
  {C.}~\bibnamefont {Zhou}}, \bibinfo {author} {\bibfnamefont {R.}~\bibnamefont
  {Li}}, \bibinfo {author} {\bibfnamefont {Y.}~\bibnamefont {Jia}}, \bibinfo
  {author} {\bibfnamefont {M.}~\bibnamefont {Liu}}, \bibinfo {author}
  {\bibfnamefont {L.}~\bibnamefont {Frunzio}}, \ and\ \bibinfo {author}
  {\bibfnamefont {R.~J.}\ \bibnamefont {Schoelkopf}},\ }\href {\doibase
  10.1038/s41467-024-47857-6} {\bibfield  {journal} {\bibinfo  {journal}
  {Nature Communications}\ }\textbf {\bibinfo {volume} {15}},\ \bibinfo {pages}
  {3687} (\bibinfo {year} {2024})}\BibitemShut {NoStop}%
\bibitem [{\citenamefont {Wenner}\ \emph {et~al.}(2011)\citenamefont {Wenner},
  \citenamefont {Barends}, \citenamefont {Bialczak}, \citenamefont {Chen},
  \citenamefont {Kelly}, \citenamefont {Lucero}, \citenamefont {Mariantoni},
  \citenamefont {Megrant}, \citenamefont {O{\textquoteright}Malley},
  \citenamefont {Sank}, \citenamefont {Vainsencher}, \citenamefont {Wang},
  \citenamefont {White}, \citenamefont {Yin}, \citenamefont {Zhao},
  \citenamefont {Cleland},\ and\ \citenamefont
  {Martinis}}]{wenner_surface_2011}%
  \BibitemOpen
  \bibfield  {author} {\bibinfo {author} {\bibfnamefont {J.}~\bibnamefont
  {Wenner}}, \bibinfo {author} {\bibfnamefont {R.}~\bibnamefont {Barends}},
  \bibinfo {author} {\bibfnamefont {R.~C.}\ \bibnamefont {Bialczak}}, \bibinfo
  {author} {\bibfnamefont {Y.}~\bibnamefont {Chen}}, \bibinfo {author}
  {\bibfnamefont {J.}~\bibnamefont {Kelly}}, \bibinfo {author} {\bibfnamefont
  {E.}~\bibnamefont {Lucero}}, \bibinfo {author} {\bibfnamefont
  {M.}~\bibnamefont {Mariantoni}}, \bibinfo {author} {\bibfnamefont
  {A.}~\bibnamefont {Megrant}}, \bibinfo {author} {\bibfnamefont {P.~J.~J.}\
  \bibnamefont {O{\textquoteright}Malley}}, \bibinfo {author} {\bibfnamefont
  {D.}~\bibnamefont {Sank}}, \bibinfo {author} {\bibfnamefont {A.}~\bibnamefont
  {Vainsencher}}, \bibinfo {author} {\bibfnamefont {H.}~\bibnamefont {Wang}},
  \bibinfo {author} {\bibfnamefont {T.~C.}\ \bibnamefont {White}}, \bibinfo
  {author} {\bibfnamefont {Y.}~\bibnamefont {Yin}}, \bibinfo {author}
  {\bibfnamefont {J.}~\bibnamefont {Zhao}}, \bibinfo {author} {\bibfnamefont
  {A.~N.}\ \bibnamefont {Cleland}}, \ and\ \bibinfo {author} {\bibfnamefont
  {J.~M.}\ \bibnamefont {Martinis}},\ }\href {\doibase 10.1063/1.3637047}
  {\bibfield  {journal} {\bibinfo  {journal} {Applied Physics Letters}\
  }\textbf {\bibinfo {volume} {99}},\ \bibinfo {pages} {113513} (\bibinfo
  {year} {2011})}\BibitemShut {NoStop}%
\bibitem [{\citenamefont {Woods}\ \emph {et~al.}(2019)\citenamefont {Woods},
  \citenamefont {Calusine}, \citenamefont {Melville}, \citenamefont {Sevi},
  \citenamefont {Golden}, \citenamefont {Kim}, \citenamefont {Rosenberg},
  \citenamefont {Yoder},\ and\ \citenamefont
  {Oliver}}]{woods_determining_2019}%
  \BibitemOpen
  \bibfield  {author} {\bibinfo {author} {\bibfnamefont {W.}~\bibnamefont
  {Woods}}, \bibinfo {author} {\bibfnamefont {G.}~\bibnamefont {Calusine}},
  \bibinfo {author} {\bibfnamefont {A.}~\bibnamefont {Melville}}, \bibinfo
  {author} {\bibfnamefont {A.}~\bibnamefont {Sevi}}, \bibinfo {author}
  {\bibfnamefont {E.}~\bibnamefont {Golden}}, \bibinfo {author} {\bibfnamefont
  {D.}~\bibnamefont {Kim}}, \bibinfo {author} {\bibfnamefont {D.}~\bibnamefont
  {Rosenberg}}, \bibinfo {author} {\bibfnamefont {J.}~\bibnamefont {Yoder}}, \
  and\ \bibinfo {author} {\bibfnamefont {W.}~\bibnamefont {Oliver}},\ }\href
  {\doibase 10.1103/PhysRevApplied.12.014012} {\bibfield  {journal} {\bibinfo
  {journal} {Phys. Rev. Appl.}\ }\textbf {\bibinfo {volume} {12}},\ \bibinfo
  {pages} {014012} (\bibinfo {year} {2019})},\ \bibinfo {note} {publisher:
  American Physical Society}\BibitemShut {NoStop}%
\bibitem [{\citenamefont {{International Centre for Diffraction
  Data}}(2018)}]{ICDD_00-038-1420_TiN}%
  \BibitemOpen
  \bibfield  {author} {\bibinfo {author} {\bibnamefont {{International Centre
  for Diffraction Data}}},\ }\href@noop {} {\enquote {\bibinfo {title} {{Powder
  Diffraction File\textregistered\ Card No.\ 00-038-1420 for Titanium Nitride
  (TiN)}},}\ }\bibinfo {howpublished} {\url{https://www.icdd.com/}} (\bibinfo
  {year} {2018}),\ \bibinfo {note} {standard reference pattern for
  face-centered cubic TiN used in XRD phase identification}\BibitemShut
  {NoStop}%
\bibitem [{\citenamefont {Ohya}\ \emph {et~al.}(2013)\citenamefont {Ohya},
  \citenamefont {Chiaro}, \citenamefont {Megrant}, \citenamefont {Neill},
  \citenamefont {Barends}, \citenamefont {Chen}, \citenamefont {Kelly},
  \citenamefont {Low}, \citenamefont {Mutus}, \citenamefont {O'Malley},
  \citenamefont {Roushan}, \citenamefont {Sank}, \citenamefont {Vainsencher},
  \citenamefont {Wenner}, \citenamefont {White}, \citenamefont {Yin},
  \citenamefont {Schultz}, \citenamefont {Palmstrøm}, \citenamefont {Mazin},
  \citenamefont {Cleland},\ and\ \citenamefont
  {Martinis}}]{ohya_sputtered_2013}%
  \BibitemOpen
  \bibfield  {author} {\bibinfo {author} {\bibfnamefont {S.}~\bibnamefont
  {Ohya}}, \bibinfo {author} {\bibfnamefont {B.}~\bibnamefont {Chiaro}},
  \bibinfo {author} {\bibfnamefont {A.}~\bibnamefont {Megrant}}, \bibinfo
  {author} {\bibfnamefont {C.}~\bibnamefont {Neill}}, \bibinfo {author}
  {\bibfnamefont {R.}~\bibnamefont {Barends}}, \bibinfo {author} {\bibfnamefont
  {Y.}~\bibnamefont {Chen}}, \bibinfo {author} {\bibfnamefont {J.}~\bibnamefont
  {Kelly}}, \bibinfo {author} {\bibfnamefont {D.}~\bibnamefont {Low}}, \bibinfo
  {author} {\bibfnamefont {J.}~\bibnamefont {Mutus}}, \bibinfo {author}
  {\bibfnamefont {P.}~\bibnamefont {O'Malley}}, \bibinfo {author}
  {\bibfnamefont {P.}~\bibnamefont {Roushan}}, \bibinfo {author} {\bibfnamefont
  {D.}~\bibnamefont {Sank}}, \bibinfo {author} {\bibfnamefont {A.}~\bibnamefont
  {Vainsencher}}, \bibinfo {author} {\bibfnamefont {J.}~\bibnamefont {Wenner}},
  \bibinfo {author} {\bibfnamefont {T.~C.}\ \bibnamefont {White}}, \bibinfo
  {author} {\bibfnamefont {Y.}~\bibnamefont {Yin}}, \bibinfo {author}
  {\bibfnamefont {B.~D.}\ \bibnamefont {Schultz}}, \bibinfo {author}
  {\bibfnamefont {C.~J.}\ \bibnamefont {Palmstrøm}}, \bibinfo {author}
  {\bibfnamefont {B.~A.}\ \bibnamefont {Mazin}}, \bibinfo {author}
  {\bibfnamefont {A.~N.}\ \bibnamefont {Cleland}}, \ and\ \bibinfo {author}
  {\bibfnamefont {J.~M.}\ \bibnamefont {Martinis}},\ }\href
  {https://arxiv.org/abs/1306.2966} {\enquote {\bibinfo {title} {Sputtered
  {T}i{N} films for superconducting coplanar waveguide resonators},}\ }
  (\bibinfo {year} {2013}),\ \Eprint {http://arxiv.org/abs/1306.2966}
  {arXiv:1306.2966 [cond-mat.supr-con]} \BibitemShut {NoStop}%
\bibitem [{\citenamefont {Zoestbergen}(2000)}]{Zoestbergen_Xray_2000}%
  \BibitemOpen
  \bibfield  {author} {\bibinfo {author} {\bibfnamefont {E.}~\bibnamefont
  {Zoestbergen}},\ }\emph {\bibinfo {title} {X-ray analysis of protective
  coatings}},\ \href@noop {} {Ph.D. thesis},\ \bibinfo  {school} {University of
  Groningen} (\bibinfo {year} {2000})\BibitemShut {NoStop}%
\bibitem [{\citenamefont {Chen}\ and\ \citenamefont
  {Lu}(2005)}]{chen_oxidation_2005}%
  \BibitemOpen
  \bibfield  {author} {\bibinfo {author} {\bibfnamefont {H.-Y.}\ \bibnamefont
  {Chen}}\ and\ \bibinfo {author} {\bibfnamefont {F.-H.}\ \bibnamefont {Lu}},\
  }\href {\doibase 10.1116/1.1914815} {\bibfield  {journal} {\bibinfo
  {journal} {Journal of Vacuum Science \& Technology A: Vacuum, Surfaces, and
  Films}\ }\textbf {\bibinfo {volume} {23}},\ \bibinfo {pages} {1006} (\bibinfo
  {year} {2005})}\BibitemShut {NoStop}%
\bibitem [{\citenamefont {Rumble}(2022)}]{Rumble_CRC_2022}%
  \BibitemOpen
  \bibinfo {editor} {\bibfnamefont {J.~R.}\ \bibnamefont {Rumble}},\ ed.,\
  \href@noop {} {\emph {\bibinfo {title} {CRC Handbook of Chemistry and
  Physics}}},\ \bibinfo {edition} {103rd}\ ed.\ (\bibinfo  {publisher} {CRC
  Press},\ \bibinfo {year} {2022})\ \bibinfo {note} {internet version 2021 =
  103rd print edition}\BibitemShut {NoStop}%
\bibitem [{\citenamefont {Bizn{\'a}rov{\'a}}\ \emph {et~al.}(2024)\citenamefont
  {Bizn{\'a}rov{\'a}}, \citenamefont {Osman}, \citenamefont {Rehnman},
  \citenamefont {Chayanun}, \citenamefont {Kri{\v z}an}, \citenamefont
  {Malmberg}, \citenamefont {Rommel}, \citenamefont {Warren}, \citenamefont
  {Delsing}, \citenamefont {Yurgens}, \citenamefont {Bylander},\ and\
  \citenamefont {Fadavi~Roudsari}}]{biznarova_mitigation_2024}%
  \BibitemOpen
  \bibfield  {author} {\bibinfo {author} {\bibfnamefont {J.}~\bibnamefont
  {Bizn{\'a}rov{\'a}}}, \bibinfo {author} {\bibfnamefont {A.}~\bibnamefont
  {Osman}}, \bibinfo {author} {\bibfnamefont {E.}~\bibnamefont {Rehnman}},
  \bibinfo {author} {\bibfnamefont {L.}~\bibnamefont {Chayanun}}, \bibinfo
  {author} {\bibfnamefont {C.}~\bibnamefont {Kri{\v z}an}}, \bibinfo {author}
  {\bibfnamefont {P.}~\bibnamefont {Malmberg}}, \bibinfo {author}
  {\bibfnamefont {M.}~\bibnamefont {Rommel}}, \bibinfo {author} {\bibfnamefont
  {C.}~\bibnamefont {Warren}}, \bibinfo {author} {\bibfnamefont
  {P.}~\bibnamefont {Delsing}}, \bibinfo {author} {\bibfnamefont
  {A.}~\bibnamefont {Yurgens}}, \bibinfo {author} {\bibfnamefont
  {J.}~\bibnamefont {Bylander}}, \ and\ \bibinfo {author} {\bibfnamefont
  {A.}~\bibnamefont {Fadavi~Roudsari}},\ }\href {\doibase
  10.1038/s41534-024-00868-z} {\bibfield  {journal} {\bibinfo  {journal} {npj
  Quantum Information}\ }\textbf {\bibinfo {volume} {10}},\ \bibinfo {pages}
  {78} (\bibinfo {year} {2024})}\BibitemShut {NoStop}%
\bibitem [{\citenamefont {O{\textquoteright}Connell}\ \emph
  {et~al.}(2008)\citenamefont {O{\textquoteright}Connell}, \citenamefont
  {Ansmann}, \citenamefont {Bialczak}, \citenamefont {Hofheinz}, \citenamefont
  {Katz}, \citenamefont {Lucero}, \citenamefont {McKenney}, \citenamefont
  {Neeley}, \citenamefont {Wang}, \citenamefont {Weig}, \citenamefont
  {Cleland},\ and\ \citenamefont {Martinis}}]{oconnell_microwave_2008}%
  \BibitemOpen
  \bibfield  {author} {\bibinfo {author} {\bibfnamefont {A.~D.}\ \bibnamefont
  {O{\textquoteright}Connell}}, \bibinfo {author} {\bibfnamefont
  {M.}~\bibnamefont {Ansmann}}, \bibinfo {author} {\bibfnamefont {R.~C.}\
  \bibnamefont {Bialczak}}, \bibinfo {author} {\bibfnamefont {M.}~\bibnamefont
  {Hofheinz}}, \bibinfo {author} {\bibfnamefont {N.}~\bibnamefont {Katz}},
  \bibinfo {author} {\bibfnamefont {E.}~\bibnamefont {Lucero}}, \bibinfo
  {author} {\bibfnamefont {C.}~\bibnamefont {McKenney}}, \bibinfo {author}
  {\bibfnamefont {M.}~\bibnamefont {Neeley}}, \bibinfo {author} {\bibfnamefont
  {H.}~\bibnamefont {Wang}}, \bibinfo {author} {\bibfnamefont {E.~M.}\
  \bibnamefont {Weig}}, \bibinfo {author} {\bibfnamefont {A.~N.}\ \bibnamefont
  {Cleland}}, \ and\ \bibinfo {author} {\bibfnamefont {J.~M.}\ \bibnamefont
  {Martinis}},\ }\href {\doibase 10.1063/1.2898887} {\bibfield  {journal}
  {\bibinfo  {journal} {Applied Physics Letters}\ }\textbf {\bibinfo {volume}
  {92}},\ \bibinfo {pages} {112903} (\bibinfo {year} {2008})}\BibitemShut
  {NoStop}%
\bibitem [{\citenamefont {Alto{\'e}}\ \emph {et~al.}(2022)\citenamefont
  {Alto{\'e}}, \citenamefont {Banerjee}, \citenamefont {Berk}, \citenamefont
  {Hajr}, \citenamefont {Schwartzberg}, \citenamefont {Song}, \citenamefont
  {Alghadeer}, \citenamefont {Aloni}, \citenamefont {Elowson}, \citenamefont
  {Kreikebaum}, \citenamefont {Wong}, \citenamefont {Griffin}, \citenamefont
  {Rao}, \citenamefont {Weber-Bargioni}, \citenamefont {Minor}, \citenamefont
  {Santiago}, \citenamefont {Cabrini}, \citenamefont {Siddiqi},\ and\
  \citenamefont {Ogletree}}]{altoe_localization_2022}%
  \BibitemOpen
  \bibfield  {author} {\bibinfo {author} {\bibfnamefont {M.~V.~P.}\
  \bibnamefont {Alto{\'e}}}, \bibinfo {author} {\bibfnamefont {A.}~\bibnamefont
  {Banerjee}}, \bibinfo {author} {\bibfnamefont {C.}~\bibnamefont {Berk}},
  \bibinfo {author} {\bibfnamefont {A.}~\bibnamefont {Hajr}}, \bibinfo {author}
  {\bibfnamefont {A.}~\bibnamefont {Schwartzberg}}, \bibinfo {author}
  {\bibfnamefont {C.}~\bibnamefont {Song}}, \bibinfo {author} {\bibfnamefont
  {M.}~\bibnamefont {Alghadeer}}, \bibinfo {author} {\bibfnamefont
  {S.}~\bibnamefont {Aloni}}, \bibinfo {author} {\bibfnamefont {M.~J.}\
  \bibnamefont {Elowson}}, \bibinfo {author} {\bibfnamefont {J.~M.}\
  \bibnamefont {Kreikebaum}}, \bibinfo {author} {\bibfnamefont {E.~K.}\
  \bibnamefont {Wong}}, \bibinfo {author} {\bibfnamefont {S.~M.}\ \bibnamefont
  {Griffin}}, \bibinfo {author} {\bibfnamefont {S.}~\bibnamefont {Rao}},
  \bibinfo {author} {\bibfnamefont {A.}~\bibnamefont {Weber-Bargioni}},
  \bibinfo {author} {\bibfnamefont {A.~M.}\ \bibnamefont {Minor}}, \bibinfo
  {author} {\bibfnamefont {D.~I.}\ \bibnamefont {Santiago}}, \bibinfo {author}
  {\bibfnamefont {S.}~\bibnamefont {Cabrini}}, \bibinfo {author} {\bibfnamefont
  {I.}~\bibnamefont {Siddiqi}}, \ and\ \bibinfo {author} {\bibfnamefont
  {D.~F.}\ \bibnamefont {Ogletree}},\ }\href {\doibase
  10.1103/PRXQuantum.3.020312} {\bibfield  {journal} {\bibinfo  {journal} {PRX
  Quantum}\ }\textbf {\bibinfo {volume} {3}},\ \bibinfo {pages} {020312}
  (\bibinfo {year} {2022})}\BibitemShut {NoStop}%
\bibitem [{\citenamefont {Chang}\ \emph {et~al.}(2025)\citenamefont {Chang},
  \citenamefont {Shumiya}, \citenamefont {McLellan}, \citenamefont {Zhang},
  \citenamefont {Bland}, \citenamefont {Bahrami}, \citenamefont {Mun},
  \citenamefont {Zhou}, \citenamefont {Kisslinger}, \citenamefont {Cheng},
  \citenamefont {Smitham}, \citenamefont {Pakpour-Tabrizi}, \citenamefont
  {Yao}, \citenamefont {Zhu}, \citenamefont {Liu}, \citenamefont {Cava},
  \citenamefont {Gopalakrishnan}, \citenamefont {Houck},\ and\ \citenamefont
  {de~Leon}}]{chang_eliminating_2025}%
  \BibitemOpen
  \bibfield  {author} {\bibinfo {author} {\bibfnamefont {R.~D.}\ \bibnamefont
  {Chang}}, \bibinfo {author} {\bibfnamefont {N.}~\bibnamefont {Shumiya}},
  \bibinfo {author} {\bibfnamefont {R.~A.}\ \bibnamefont {McLellan}}, \bibinfo
  {author} {\bibfnamefont {Y.}~\bibnamefont {Zhang}}, \bibinfo {author}
  {\bibfnamefont {M.~P.}\ \bibnamefont {Bland}}, \bibinfo {author}
  {\bibfnamefont {F.}~\bibnamefont {Bahrami}}, \bibinfo {author} {\bibfnamefont
  {J.}~\bibnamefont {Mun}}, \bibinfo {author} {\bibfnamefont {C.}~\bibnamefont
  {Zhou}}, \bibinfo {author} {\bibfnamefont {K.}~\bibnamefont {Kisslinger}},
  \bibinfo {author} {\bibfnamefont {G.}~\bibnamefont {Cheng}}, \bibinfo
  {author} {\bibfnamefont {B.~M.}\ \bibnamefont {Smitham}}, \bibinfo {author}
  {\bibfnamefont {A.~C.}\ \bibnamefont {Pakpour-Tabrizi}}, \bibinfo {author}
  {\bibfnamefont {N.}~\bibnamefont {Yao}}, \bibinfo {author} {\bibfnamefont
  {Y.}~\bibnamefont {Zhu}}, \bibinfo {author} {\bibfnamefont {M.}~\bibnamefont
  {Liu}}, \bibinfo {author} {\bibfnamefont {R.~J.}\ \bibnamefont {Cava}},
  \bibinfo {author} {\bibfnamefont {S.}~\bibnamefont {Gopalakrishnan}},
  \bibinfo {author} {\bibfnamefont {A.~A.}\ \bibnamefont {Houck}}, \ and\
  \bibinfo {author} {\bibfnamefont {N.~P.}\ \bibnamefont {de~Leon}},\ }\href
  {\doibase 10.1103/PhysRevLett.134.097001} {\bibfield  {journal} {\bibinfo
  {journal} {Phys. Rev. Lett.}\ }\textbf {\bibinfo {volume} {134}},\ \bibinfo
  {pages} {097001} (\bibinfo {year} {2025})},\ \bibinfo {note} {publisher:
  American Physical Society}\BibitemShut {NoStop}%
\bibitem [{\citenamefont {Verjauw}\ \emph {et~al.}(2021)\citenamefont
  {Verjauw}, \citenamefont {Poto{\v c}nik}, \citenamefont {Mongillo},
  \citenamefont {Acharya}, \citenamefont {Mohiyaddin}, \citenamefont {Simion},
  \citenamefont {Pacco}, \citenamefont {Ivanov}, \citenamefont {Wan},
  \citenamefont {Vanleenhove}, \citenamefont {Souriau}, \citenamefont {Jussot},
  \citenamefont {Thiam}, \citenamefont {Swerts}, \citenamefont {Piao},
  \citenamefont {Couet}, \citenamefont {Heyns}, \citenamefont {Govoreanu},\
  and\ \citenamefont {Radu}}]{verjauw_investigation_2021}%
  \BibitemOpen
  \bibfield  {author} {\bibinfo {author} {\bibfnamefont {J.}~\bibnamefont
  {Verjauw}}, \bibinfo {author} {\bibfnamefont {A.}~\bibnamefont {Poto{\v
  c}nik}}, \bibinfo {author} {\bibfnamefont {M.}~\bibnamefont {Mongillo}},
  \bibinfo {author} {\bibfnamefont {R.}~\bibnamefont {Acharya}}, \bibinfo
  {author} {\bibfnamefont {F.}~\bibnamefont {Mohiyaddin}}, \bibinfo {author}
  {\bibfnamefont {G.}~\bibnamefont {Simion}}, \bibinfo {author} {\bibfnamefont
  {A.}~\bibnamefont {Pacco}}, \bibinfo {author} {\bibfnamefont
  {T.}~\bibnamefont {Ivanov}}, \bibinfo {author} {\bibfnamefont
  {D.}~\bibnamefont {Wan}}, \bibinfo {author} {\bibfnamefont {A.}~\bibnamefont
  {Vanleenhove}}, \bibinfo {author} {\bibfnamefont {L.}~\bibnamefont
  {Souriau}}, \bibinfo {author} {\bibfnamefont {J.}~\bibnamefont {Jussot}},
  \bibinfo {author} {\bibfnamefont {A.}~\bibnamefont {Thiam}}, \bibinfo
  {author} {\bibfnamefont {J.}~\bibnamefont {Swerts}}, \bibinfo {author}
  {\bibfnamefont {X.}~\bibnamefont {Piao}}, \bibinfo {author} {\bibfnamefont
  {S.}~\bibnamefont {Couet}}, \bibinfo {author} {\bibfnamefont
  {M.}~\bibnamefont {Heyns}}, \bibinfo {author} {\bibfnamefont
  {B.}~\bibnamefont {Govoreanu}}, \ and\ \bibinfo {author} {\bibfnamefont
  {I.}~\bibnamefont {Radu}},\ }\href {\doibase
  10.1103/PhysRevApplied.16.014018} {\bibfield  {journal} {\bibinfo  {journal}
  {Phys. Rev. Appl.}\ }\textbf {\bibinfo {volume} {16}},\ \bibinfo {pages}
  {014018} (\bibinfo {year} {2021})},\ \bibinfo {note} {publisher: American
  Physical Society}\BibitemShut {NoStop}%
\bibitem [{\citenamefont {Gupta}\ \emph {et~al.}(2026)\citenamefont {Gupta},
  \citenamefont {Pereira}, \citenamefont {Koch}, \citenamefont {Bruckmoser},
  \citenamefont {Singer}, \citenamefont {Schoof}, \citenamefont {Kompatscher},
  \citenamefont {Filipp},\ and\ \citenamefont {Tornow}}]{gupta_high_2026}%
  \BibitemOpen
  \bibfield  {author} {\bibinfo {author} {\bibfnamefont {H.}~\bibnamefont
  {Gupta}}, \bibinfo {author} {\bibfnamefont {R.}~\bibnamefont {Pereira}},
  \bibinfo {author} {\bibfnamefont {L.}~\bibnamefont {Koch}}, \bibinfo {author}
  {\bibfnamefont {N.}~\bibnamefont {Bruckmoser}}, \bibinfo {author}
  {\bibfnamefont {M.}~\bibnamefont {Singer}}, \bibinfo {author} {\bibfnamefont
  {B.}~\bibnamefont {Schoof}}, \bibinfo {author} {\bibfnamefont
  {M.}~\bibnamefont {Kompatscher}}, \bibinfo {author} {\bibfnamefont
  {S.}~\bibnamefont {Filipp}}, \ and\ \bibinfo {author} {\bibfnamefont
  {M.}~\bibnamefont {Tornow}},\ }\href {\doibase 10.1038/s43246-025-01068-8}
  {\bibfield  {journal} {\bibinfo  {journal} {Communications Materials}\
  }\textbf {\bibinfo {volume} {7}},\ \bibinfo {pages} {58} (\bibinfo {year}
  {2026})}\BibitemShut {NoStop}%
\bibitem [{\citenamefont {Quintana}\ \emph {et~al.}(2014)\citenamefont
  {Quintana}, \citenamefont {Megrant}, \citenamefont {Chen}, \citenamefont
  {Dunsworth}, \citenamefont {Chiaro}, \citenamefont {Barends}, \citenamefont
  {Campbell}, \citenamefont {Chen}, \citenamefont {Hoi}, \citenamefont
  {Jeffrey}, \citenamefont {Kelly}, \citenamefont {Mutus}, \citenamefont
  {O'Malley}, \citenamefont {Neill}, \citenamefont {Roushan}, \citenamefont
  {Sank}, \citenamefont {Vainsencher}, \citenamefont {Wenner}, \citenamefont
  {White}, \citenamefont {Cleland},\ and\ \citenamefont
  {Martinis}}]{quintana_characterization_2014}%
  \BibitemOpen
  \bibfield  {author} {\bibinfo {author} {\bibfnamefont {C.~M.}\ \bibnamefont
  {Quintana}}, \bibinfo {author} {\bibfnamefont {A.}~\bibnamefont {Megrant}},
  \bibinfo {author} {\bibfnamefont {Z.}~\bibnamefont {Chen}}, \bibinfo {author}
  {\bibfnamefont {A.}~\bibnamefont {Dunsworth}}, \bibinfo {author}
  {\bibfnamefont {B.}~\bibnamefont {Chiaro}}, \bibinfo {author} {\bibfnamefont
  {R.}~\bibnamefont {Barends}}, \bibinfo {author} {\bibfnamefont
  {B.}~\bibnamefont {Campbell}}, \bibinfo {author} {\bibfnamefont
  {Y.}~\bibnamefont {Chen}}, \bibinfo {author} {\bibfnamefont {I.-C.}\
  \bibnamefont {Hoi}}, \bibinfo {author} {\bibfnamefont {E.}~\bibnamefont
  {Jeffrey}}, \bibinfo {author} {\bibfnamefont {J.}~\bibnamefont {Kelly}},
  \bibinfo {author} {\bibfnamefont {J.~Y.}\ \bibnamefont {Mutus}}, \bibinfo
  {author} {\bibfnamefont {P.~J.~J.}\ \bibnamefont {O'Malley}}, \bibinfo
  {author} {\bibfnamefont {C.}~\bibnamefont {Neill}}, \bibinfo {author}
  {\bibfnamefont {P.}~\bibnamefont {Roushan}}, \bibinfo {author} {\bibfnamefont
  {D.}~\bibnamefont {Sank}}, \bibinfo {author} {\bibfnamefont {A.}~\bibnamefont
  {Vainsencher}}, \bibinfo {author} {\bibfnamefont {J.}~\bibnamefont {Wenner}},
  \bibinfo {author} {\bibfnamefont {T.~C.}\ \bibnamefont {White}}, \bibinfo
  {author} {\bibfnamefont {A.~N.}\ \bibnamefont {Cleland}}, \ and\ \bibinfo
  {author} {\bibfnamefont {J.~M.}\ \bibnamefont {Martinis}},\ }\href {\doibase
  10.1063/1.4893297} {\bibfield  {journal} {\bibinfo  {journal} {Applied
  Physics Letters}\ }\textbf {\bibinfo {volume} {105}},\ \bibinfo {pages}
  {062601} (\bibinfo {year} {2014})}\BibitemShut {NoStop}%
\bibitem [{\citenamefont {Dunsworth}\ \emph {et~al.}(2017)\citenamefont
  {Dunsworth}, \citenamefont {Megrant}, \citenamefont {Quintana}, \citenamefont
  {Chen}, \citenamefont {Barends}, \citenamefont {Burkett}, \citenamefont
  {Foxen}, \citenamefont {Chen}, \citenamefont {Chiaro}, \citenamefont
  {Fowler}, \citenamefont {Graff}, \citenamefont {Jeffrey}, \citenamefont
  {Kelly}, \citenamefont {Lucero}, \citenamefont {Mutus}, \citenamefont
  {Neeley}, \citenamefont {Neill}, \citenamefont {Roushan}, \citenamefont
  {Sank}, \citenamefont {Vainsencher}, \citenamefont {Wenner}, \citenamefont
  {White},\ and\ \citenamefont {Martinis}}]{dunsworth_characterization_2017}%
  \BibitemOpen
  \bibfield  {author} {\bibinfo {author} {\bibfnamefont {A.}~\bibnamefont
  {Dunsworth}}, \bibinfo {author} {\bibfnamefont {A.}~\bibnamefont {Megrant}},
  \bibinfo {author} {\bibfnamefont {C.}~\bibnamefont {Quintana}}, \bibinfo
  {author} {\bibfnamefont {Z.}~\bibnamefont {Chen}}, \bibinfo {author}
  {\bibfnamefont {R.}~\bibnamefont {Barends}}, \bibinfo {author} {\bibfnamefont
  {B.}~\bibnamefont {Burkett}}, \bibinfo {author} {\bibfnamefont
  {B.}~\bibnamefont {Foxen}}, \bibinfo {author} {\bibfnamefont
  {Y.}~\bibnamefont {Chen}}, \bibinfo {author} {\bibfnamefont {B.}~\bibnamefont
  {Chiaro}}, \bibinfo {author} {\bibfnamefont {A.}~\bibnamefont {Fowler}},
  \bibinfo {author} {\bibfnamefont {R.}~\bibnamefont {Graff}}, \bibinfo
  {author} {\bibfnamefont {E.}~\bibnamefont {Jeffrey}}, \bibinfo {author}
  {\bibfnamefont {J.}~\bibnamefont {Kelly}}, \bibinfo {author} {\bibfnamefont
  {E.}~\bibnamefont {Lucero}}, \bibinfo {author} {\bibfnamefont {J.~Y.}\
  \bibnamefont {Mutus}}, \bibinfo {author} {\bibfnamefont {M.}~\bibnamefont
  {Neeley}}, \bibinfo {author} {\bibfnamefont {C.}~\bibnamefont {Neill}},
  \bibinfo {author} {\bibfnamefont {P.}~\bibnamefont {Roushan}}, \bibinfo
  {author} {\bibfnamefont {D.}~\bibnamefont {Sank}}, \bibinfo {author}
  {\bibfnamefont {A.}~\bibnamefont {Vainsencher}}, \bibinfo {author}
  {\bibfnamefont {J.}~\bibnamefont {Wenner}}, \bibinfo {author} {\bibfnamefont
  {T.~C.}\ \bibnamefont {White}}, \ and\ \bibinfo {author} {\bibfnamefont
  {J.~M.}\ \bibnamefont {Martinis}},\ }\href {\doibase 10.1063/1.4993577}
  {\bibfield  {journal} {\bibinfo  {journal} {Applied Physics Letters}\
  }\textbf {\bibinfo {volume} {111}},\ \bibinfo {pages} {022601} (\bibinfo
  {year} {2017})}\BibitemShut {NoStop}%
\bibitem [{\citenamefont {Murray}(2021)}]{murray_material_2021}%
  \BibitemOpen
  \bibfield  {author} {\bibinfo {author} {\bibfnamefont {C.~E.}\ \bibnamefont
  {Murray}},\ }\href {\doibase 10.1016/j.mser.2021.100646} {\bibfield
  {journal} {\bibinfo  {journal} {Materials Science and Engineering: R:
  Reports}\ }\textbf {\bibinfo {volume} {146}},\ \bibinfo {pages} {100646}
  (\bibinfo {year} {2021})}\BibitemShut {NoStop}%
\end{thebibliography}%

\end{document}